\documentclass[printer]{aa}

\usepackage{newtxtext,newtxmath}

\usepackage[T1]{fontenc}
\usepackage[normalem]{ulem}

\DeclareRobustCommand{\VAN}[3]{#2}
\let\VANthebibliography\thebibliography
\def\thebibliography{\DeclareRobustCommand{\VAN}[3]{##3}\VANthebibliography}

\usepackage{graphicx}	
\usepackage{amsmath}	
\usepackage{amssymb}	
\usepackage{color}
\usepackage{booktabs}
\usepackage{xspace}
\usepackage{bm}
\usepackage{verbatim}
\usepackage{float}
\usepackage{tensor}
\usepackage{xcolor}
\usepackage[colorlinks]{hyperref}
\hypersetup{
    colorlinks=true,
    linkbordercolor={white},
    linkcolor={blue},
    citecolor={blue}
}

\newcommand{\lpnofont}{$\widetilde{\text{LP}}$\xspace}
\newcommand{\cpnofont}{$\widetilde{\text{CP}}$\xspace}
\newcommand{\lp}{{\fontsize{9pt}{9pt}\selectfont$\widetilde{\text{LP}}$}\xspace}
\newcommand{\cp}{{\fontsize{9pt}{9pt}\selectfont$\widetilde{\text{CP}}$}\xspace}
\newcommand{\quloop}{$\mathcal{Q}-\mathcal{U}$ {\rm loop}\xspace}
\newcommand{\quloops}{$\mathcal{Q}-\mathcal{U}$ {\rm loops}\xspace}
\newcommand{\qu}{$\mathcal{Q}-\mathcal{U}$\xspace}
\newcommand{\rg}{$r_{\rm g}$\xspace}

\newcommand\monika[1]{{\color{red}[MM: #1]}}

\newcommand\mw[1]{{\color{magenta}[MW: #1]}}

\title{Polarimetric signatures of hot spots in black hole accretion flows}

\author{J. Vos\inst{\ref{inst1}}\thanks{\email{jt.vos@astro.ru.nl}} \and M. Mo{\'s}cibrodzka\inst{\ref{inst1}} \and
	M. Wielgus\inst{\ref{inst2}} 
}

\institute{
Department of Astrophysics/IMAPP, Radboud University, PO Box 9010, 6500 GL Nijmegen, the Netherlands\label{inst1} \and
Max-Planck-Institut f\"ur Radioastronomie, Auf dem H\"ugel 69, D-53121 Bonn, Germany\label{inst2}
}

\date{Received <date> / Accepted <date>}

\abstract{The flaring events observed in the Sagittarius A* supermassive black hole system can be attributed to the non-homogeneous nature of the near-horizon accretion flow. Bright regions in this flow may be associated with density or temperature anisotropies, so-called ``bright spot'' or ``hot spots''. Such orbiting features may explain observations at infrared wavelengths as well as recent findings at millimeter wavelengths.
}
{In this work, we study the emission from an orbiting equatorial bright spot, imposed on a radiatively inefficient accretion flow background, to find polarimetric features indicative of the underlying magnetic field structure and other system variables including inclination angle, spot size, black hole spin, and more. Specifically, we investigate the impact of these parameters on the Stokes \qu 
signatures that commonly exhibit a typical double loop (pretzel-like) structure.}
{Our semi-analytical model, describing the underlying plasma conditions and the orbiting spot, is built within the framework of the numerical radiative transfer code {\tt ipole}, which calculates synchroton emission at 230 GHz.}
{We showcase the wide variety of \quloop signatures and the relation between inner and outer loops. For the \emph{vertical} magnetic field topology, the inner \quloop is explained by the suppression of the synchrotron emission as seen by the distant observer. For the radial and toroidal magnetic field topologies, the inner \quloop corresponds to the part of the orbit where the spot it is receding with respect to the observer. }
{Based on our models we conclude that it is possible to constrain the underlying magnetic field topology with an analysis of the \quloop geometry, particularly in combination with a circular polarization measurements.}

\keywords{black hole physics -- relativistic processes -- radiative transfer -- methods: numerical}

\begin{document}


\maketitle

\section{Introduction}
The supermassive black hole (SMBH/BH) at the center of the Milky Way, Sagittarius A* (hereafter Sgr~A*), is surrounded by a radiatevely inefficient accretion flow (RIAF) with  bolometric luminosity at a strongly sub-Eddington level ($L_{\rm bol}\approx 10^{-10} L_{\rm Edd}$, where $L_{\rm Edd}$ is the Eddington luminosity; for reviews of Sgr~A* observational properties see, e.g.,~\citealt{genzel10,morris12}).
The environment surrounding SMBH is emitting photons across the entire electromagnetic spectrum from radio to $\gamma-$rays, but the source spectral energy distribution peaks in the millimeter band. 
The millimeter band emission is dominated by the synchrotron process occurring in the immediate vicinity of the SMBH \citep[e.g.,][]{narayan95,SgrAPaperI,SgrAPaperV}.  
Sgr~A* is known to display flaring behaviour on the top of the base "quiescent" stochastic variability. During these flaring events, observed on a daily basis, the signal can increase by up to $\sim 100$ times the mean quiescent flux density in the near-infrared (NIR) \citep[e.g.,][]{dodds-eden09,witzel18,witzel21} and X-ray \citep[e.g.,][]{porquet08,nowak12,do19}.
The flux density is much larger at millimeter to radio wavelengths than the mean NIR and X-ray values and displays only marginal flaring behavior \citep[e.g.,][]{SgraP2, wielgus22}. This apparent discrepancy is attributed to effects associated with particle heating/acceleration occurring near the SMBH but the details of such flaring sites such as their location, geometry, magnetization, and evolution are currently not well-understood \citep[and references therein]{witzel18}.

One of the proposed models for explaining the flaring activity of Sgr~A* is the formation of plasmoids\footnote{A number of other non-plasmoid explanations \citep[see, e.g.,][]{witzel21} have also been proposed to explain the NIR variability.} near the BH event horizon.
The term "plasmoid" is broadly used, but we employ the term to describe a "localized energized accretion flow feature". These bright blob-like features are also referred to as "hot spots" or "bright spots". 
Such a scenario has been adopted to explain NIR flares observed using the GRAVITY instrument \citep{gravity_plasmoids_18,gravity_jimenez20} and polarimetric variability at millimeter wavelengths \citep{wielgus22polar} observed with the Atacama Large Millimeter/submillimeter Array (ALMA).

It is still largely unknown how a plasmoid would observationally manifest itself at millimeter wavelengths. With the Event Horizon Telescope \citep[EHT;][]{eht1,eht2} it has recently become possible to reach a sufficient angular resolution to image event horizon scale features of Sgr~A*, both in total intensity \citep{SgrAPaperI,SgrAPaperIII} and in linear polarization \citep[so far only published for the M87* SMBH;][]{eht7,eht8}. While there is no unambiguous trace of orbiting hot spots in the EHT images of Sgr~A* published so far, we turn to investigating their impact on the millimeter light curves. In particular, Sgr~A*’s linear polarization observations (during flares) in both the NIR and millimeter bands display similar looping behavior in the Stokes ${\mathcal Q}-{\mathcal U}$ diagrams \citep{marrone06phd,gravity_jimenez20,wielgus22polar}, suggesting that the flaring events should manifest themselves in the millimeter band light curves of the unresolved compact source.
Additionally, in the millimeter band we observe persistent net circular polarization (Stokes~$\mathcal{V}$) of the Sgr~A* light curves at the level of $\sim-1\%$ \citep{munoz12,goddi21,wielgus22polar}. 
Circular polarization measurements obtained during a ${\mathcal Q}-{\mathcal U}$ loop may help us to further constrain the physics of the flares. 

On the theoretical front there has been a significant effort to model Sgr~A* flares with plasmoids. Plasmoids emission or their formation are currently modeled using a few distinct approaches: (i) semi-analytic geometric models \citep{gelles21}, (ii) semi-analytic accretion flow models with a Gaussian fluctuation in it \citep{broderick05,broderick06, gravity_jimenez20}, (iii) general relativistic magnetohydrodynamic (MHD/GRMHD) simulations \citep[see, e.g.,][]{murphy13,ripperda22} and (iv) general relativistic particle-in-cell (PIC/GRPIC) simulations \citep[see, e.g.,][]{cerutti14a,sironi16,crinquand21}. The approaches (ii), (iii) and (iv) include all general relativistic effects.
Approaches (i) and (ii) focus on predicting radiative properties of plasmoids; approaches (iii) and (iv) mostly aim at following plasmoid formation and dynamics.
In approaches (iii) and (iv) plasmoids are associated with magnetic reconnection events that heat the plasma \citep[e.g.,][]{rowan17}.
Under certain conditions the reconnection current sheets break into a chain of plasmoids which then grow by merging with other plasmoids and evolve further \citep{loureiro12}. Plasmoids may therefore observationally appear as hot spots.

The main advantage of semi-analytic models of plasmoids (approach (i) and (ii)) is that their observational appearance can be accurately predicted using first-principle radiative transfer calculations.  
The first semi-analytical radiative models of plasmoid were developed in \citet{broderick05}, where the authors investigated the total intensity and polarized light curves and images of an optically thick sphere that orbits the BH in the equatorial plane. 
In \citet{broderick06}, the previous models are expanded by adding a RIAF background emission and a different underlying magnetic field. 
Other notable papers that discuss radiation from (semi-analytical) Gaussian structures include: \citet[][]{hamaus09,Zamaninasab2010,broderick11b,broderick11a,pu16,tiede20}.
Another class of these models discusses the hot spot's formation and evolution along the jet of the BH \citep{younsi15,ball21}.

The main focus of this work is to investigate the Stokes $\mathcal{I}$, $\mathcal{Q}$, $\mathcal{U}$, and $\mathcal{V}$ properties of the semi-analytic Kerr black hole RIAF + hot spot model, and their dependence on the model parameters, particularly the magnetic field topology.
We argue that the polarimetric features introduced by an equatorially orbiting hot spot, in addition to a RIAF background, are indicative of the underlying magnetic field geometry and spacetime. 
More specifically, we demonstrate the utility of Stokes $\mathcal{Q}-\mathcal{U}$ diagrams (first shown in \citealt{marrone06phd}; also discussed in \citealt{gravity_jimenez20,gelles21}) for differentiating between a wide variety of simulated scenarios that can help constrain black hole spin, inclination, and orbital velocity of the entire system.
Additionally, we discuss how circular polarization from orbiting plasmoids can help differentiate between dominant magnetic field geometries.
While the NIR emission is well explained by an orbiting hot spot alone, the millimeter emission is expected to be dominated by the accretion disk background.
The interplay between background and orbiting hot spot is another problem of interest that is investigated in this work.

Our work is motivated by the new full Stokes, high-sensitivity millimeter light curve datasets available for Sgr~A* \citep[][]{wielgus22,wielgus22polar}, and in particular by the identification of orbiting hot spot signatures in ALMA 230 GHz observations \citep{wielgus22polar}. The paper is structured as follows.
We describe our model and its parameters in Sect.~\ref{sec:model}. 
In Sect.~\ref{sec:results} we report 
the polarimetric properties of hot spots as a function of the model free parameters. We focus on discussing polarimetric light curves, \quloops, and circular polarization. 
We discuss our results and compare them to previous works in Sect.~\ref{sec:discussion}. 
Finally, we summarize our main findings and conclude in Sect.~\ref{sec:summary}. 

\section{Model}\label{sec:model}

\subsection{Numerical setup}

Our semi-analytic model of the accretion flow with an orbiting bright spot around a Kerr black hole is built within {\tt ipole} - a ray-tracing code for covariant general relativistic transport of polarized light in curved spacetimes. The detailed description of the numerical scheme is presented by \citet{moscibrodzka18}. 
The radiative transfer model requires defining the underlying plasma density, electron temperature, and magnetic field strength and geometry to calculate thermal synchrotron emission.
Given a plasma configuration, {\tt ipole} generates maps of Stokes $\mathcal{I}$, $\mathcal{Q}$, $\mathcal{U}$, $\mathcal{V}$ at chosen observing frequencies including emission, self-absorption and Faraday effects (rotation and conversion). In this work, we will limit ourselves to model thermal synchrotron emission at 230 GHz ($\lambda=1.3$\,mm, i.e. millimeter band) with a standard image resolution of $256 \times 256$ (pixels). The images are pixel integrated to generate light curves. The radiative transfer model takes into account the finite speed of light which indicates a "slow light" approach.

All calculations are carried out using the Kerr metric described by the Kerr-Schild coordinate system ($t,r,\theta,\phi$) \citep[which is well-described in, e.g.,][]{mks2004} with the standard metric signature (- + + +).
In the following sections, we measure distances in units of gravitational radii ($r_g = GM/c^2$, where $G$ and $c$ are the gravitational constant and the speed of light, respectively). Time is measured in units of $r_g/c$.

\subsection{Background accretion flow}
Our background accretion flow model is inspired by a radiatively inefficient accretion disk presented in \citet{broderick06} (see also \citealt{fraga16}). At the equator, we assume that the accretion flow around the Kerr black hole is Keplerian. The dimensionless black hole spin parameter is defined as $a_* \equiv Jc/GM^2$, where $J$ is the black hole angular momentum. In this work, we consider $a_* =0,\pm0.5,\pm0.9375$.
The various quantities describing the disk properties, with the subscript $b$ for background, are denoted by:
\begin{align}
    &n_{e,b} = n_{0}\,\left(\frac{r}{r_g}\right)^{-3/2}\,\exp \left( -\tfrac{\cos^2(\theta)}{2\sigma^2} \right) \label{eq:electron_density} , \\
    &\Theta_{e,b} = \Theta_{0} \left( \frac{r}{r_g} \right)^{-0.84} \, , \\
    &B = B_0 \left(\frac{r}{r_g} \right)^{-1} .
\end{align}
Here, $n_e$ is the electron density, $\Theta_e\equiv k_B T_e /m_e c^2$ is the dimensionless electron temperature, $B$ is the magnetic field strength, and $\sigma$ parameter characterises the accretion disk vertical thickness.

All user-defined scaling factors and variables (including $n_0$ and $\Theta_0$) are listed in Table~\ref{tab:base_model_param}.
Note that we do not fit these parameters to any existing observations of Sgr~A* in this work, because the source is variable and our static background model is not.
Nevertheless, we gauge the model parameters so that the radial, $i=30^\circ$ case approximately matches the observed flux density of Sgr~A* at 230\,GHz, which is $\sim$2.5\,Jy \citep{johnson15,bower19,wielgus22}.

The electron density $n_e$ is set to a small value for radii smaller than the innermost stable circular orbit \citep[ISCO, see][]{bardeen72}. Therefore, emission from within this region is minimal except from the photon ring in certain configurations.
Additionally, as the ISCO is an spin-dependent quantity (see Table~\ref{tab:periods}) that becomes smaller for increasing $a_*$, we find that the overall $n_{e,b}$ is higher for the positive spin cases. 
Therefore, as the spot's $n_e$ (see next subsection) is unchanged for all cases, the background emission will be more pronounced in the positive spin ($a_*>0$) than in the negative spin ($a_*<0$) cases.

What is the model's magnetic field geometry?
We investigate three basic magnetic field topologies denoted simply as \emph{radial}, \emph{toroidal}, \emph{vertical}.
In purely poloidal cases (radial and vertical topology), the magnetic fields, in a given coordinate frame, are described with the azimuthal component of the vector potential. 
For the radial (monopolar) magnetic field this is;
\begin{equation}
    A_\phi = 1 - \cos(\theta),
\end{equation}
and for the vertical magnetic field this is;
\begin{equation}
    A_\phi = r \sin(\theta).
\end{equation}
While the magnetic fields of the former cases are defined through the vector potential, the toroidal magnetic field is straightforwardly defined by 
\begin{equation}
    B_\phi = 1.
\end{equation}

\subsection{Orbiting spot}

In addition to the background, we introduce a spot in a circular orbit in the equatorial plane at a fixed distance to the SMBH.
The spot's location is denoted, in Kerr-Schild coordinates ($t,r,\theta,\phi$), by:
\begin{align}
    &x_s^0 = t \, , \\
    &x_s^1 = r_\textrm{spot} = 11 \, r_\text{g} \, , \\
    &x_s^2 = \theta_\textrm{spot} = \pi/2 \, , \\
    &x_s^3 = \phi_\textrm{spot} = 2\pi t/P. 
\end{align}
Here, $t$ is the time coordinate along the null-geodesic, which implies that the spot's position is not fixed as a light ray moves through it.
The radius at which the spot orbits, $r_\textrm{spot}$, is fixed for the majority of this work.
Only the $\phi_\textrm{spot}$ coordinate changes over time and is determined by the orbital period $P$ (see Table~\ref{tab:periods}). We typically assume Keplerian rotation, that is
\begin{equation}
    P = 2\pi \left( r^{3/2} + a_* \right) \, [r_\mathrm{g}/c],
\end{equation}
and that the spot co-rotates with the background RIAF. For observers with viewing angles $i<90^\circ$ ($i>90^\circ$) the spot orbits the SMBH counterclockwise (clockwise).

Our implementation of a bright spot can be thought of as an over-density with a higher temperature, which effectively makes the spot brighter than the background emission.
The spot's electron density $n_{e,s}$ and dimensionless electron temperature $\Theta_{e,s}$ are defined as:
\begin{align}
    &n_{e,s} = \tfrac{n_0}{3} \exp \left( - \frac{\Delta x^k \Delta x_k}{2R_s^2} \right) \, ,\\    
    &\Theta_{e, s} = \Theta_0 \left( \frac{r}{r_g} \right)^{-0.84} \exp \left( - \frac{\Delta x^k \Delta x_k}{2R_s^2} \right),
\end{align}
where $\Delta x^k \equiv x^k - x_s^k$, with $k \in \{ 1,2,3 \}$, is the Euclidean distance from the spot center ($x_s^k$) to the photon (on the null-geodesic) which eventually falls on the camera.
Note that in our model the spot continues its orbit while the radiative transfer equations are evolved (a.k.a. ``slow light'').
The entire, disk + spot, system is then simply described by;
\begin{align}
    &n_e = n_{e,b} + n_{e,s} \, , \\
    &\Theta_e = \Theta_{e,b} + \Theta_{e,s}.
\end{align}

The characteristic radius of the Gaussian hot spot is fixed to $R_s = 1.5$ \rg, which makes it consistent with the assumptions of \citet{broderick05,broderick06}.
Additionally, we enforce a sharp spot boundary by only including matter if $n_{e,s}>10^5$ cm$^{-3}$.
Compared with plasmoids formed in recent GRMHD studies \citep{ripperda20,bransgrove21}, our spot's size is large but still possible. On the other hand, simulations of \citet{ripperda22} indicate formation of flux tubes that could manifest as observable orbiting hot spots, and are generally larger than the size that we assume.

\begin{figure*}
    \centering
    \includegraphics[width=\textwidth]{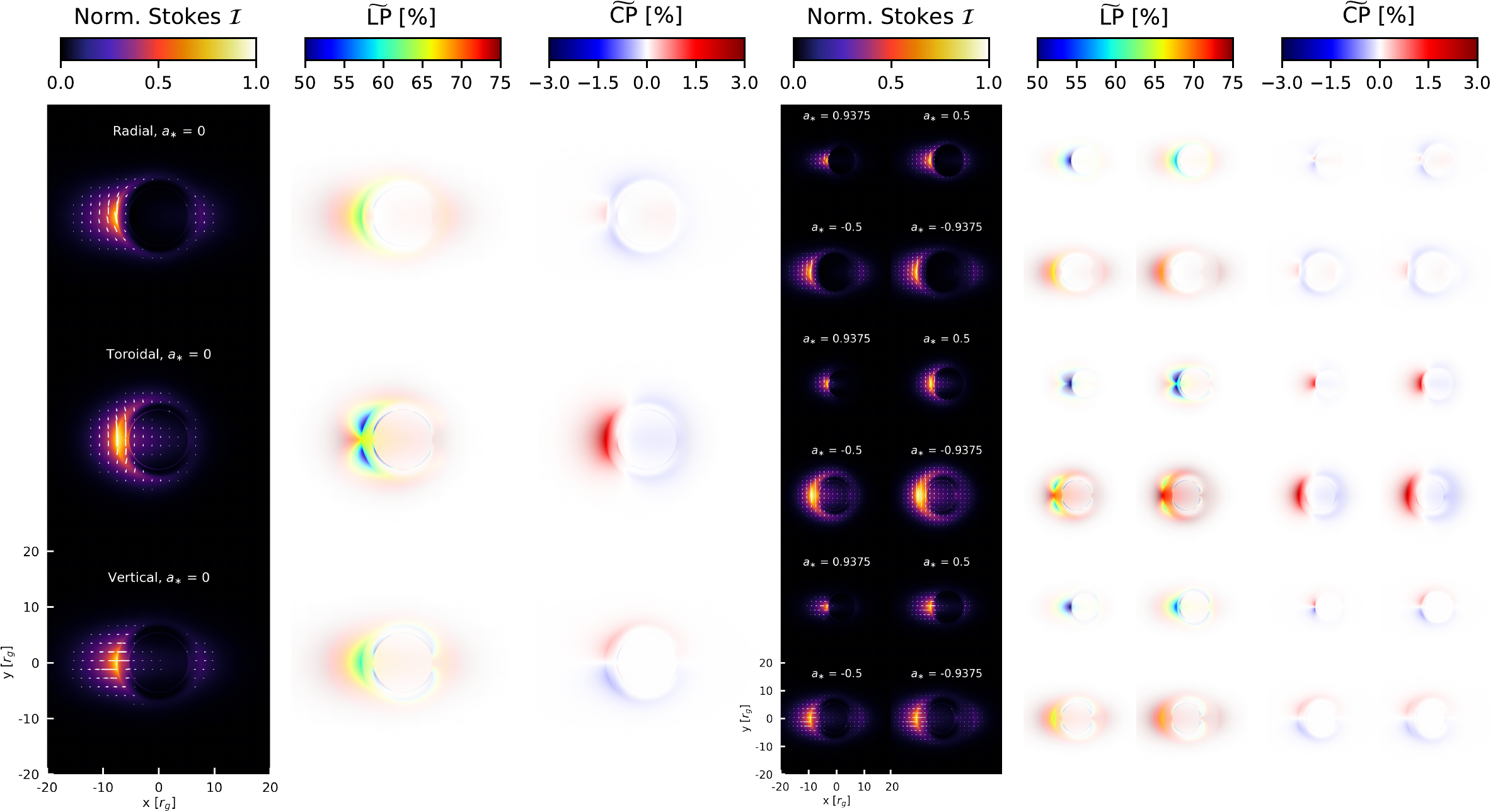}
    \caption{
    An edge-on view ($i = 90^\circ$) of the disk background in the radial (\emph{top row}), toroidal (\emph{middle row}) and vertical (\emph{bottom row}) magnetic field configurations for a field of view of $40 \, r_g \times 40 \, r_g$. The \emph{left panel} displays the zero-spin cases ($a_* = 0$), while the \emph{right panel} gives an overview of the other spin-cases. The white ticks, which are plotted over the Stokes $\mathcal{I}$ maps, denote the EVPA scaled according to {\fontsize{8pt}{8pt}\selectfont\lpnofont}.
    }
    \label{fig:disk_i90_a0}
\end{figure*}

\begin{figure*}
    \centering
    \includegraphics[width=\textwidth]{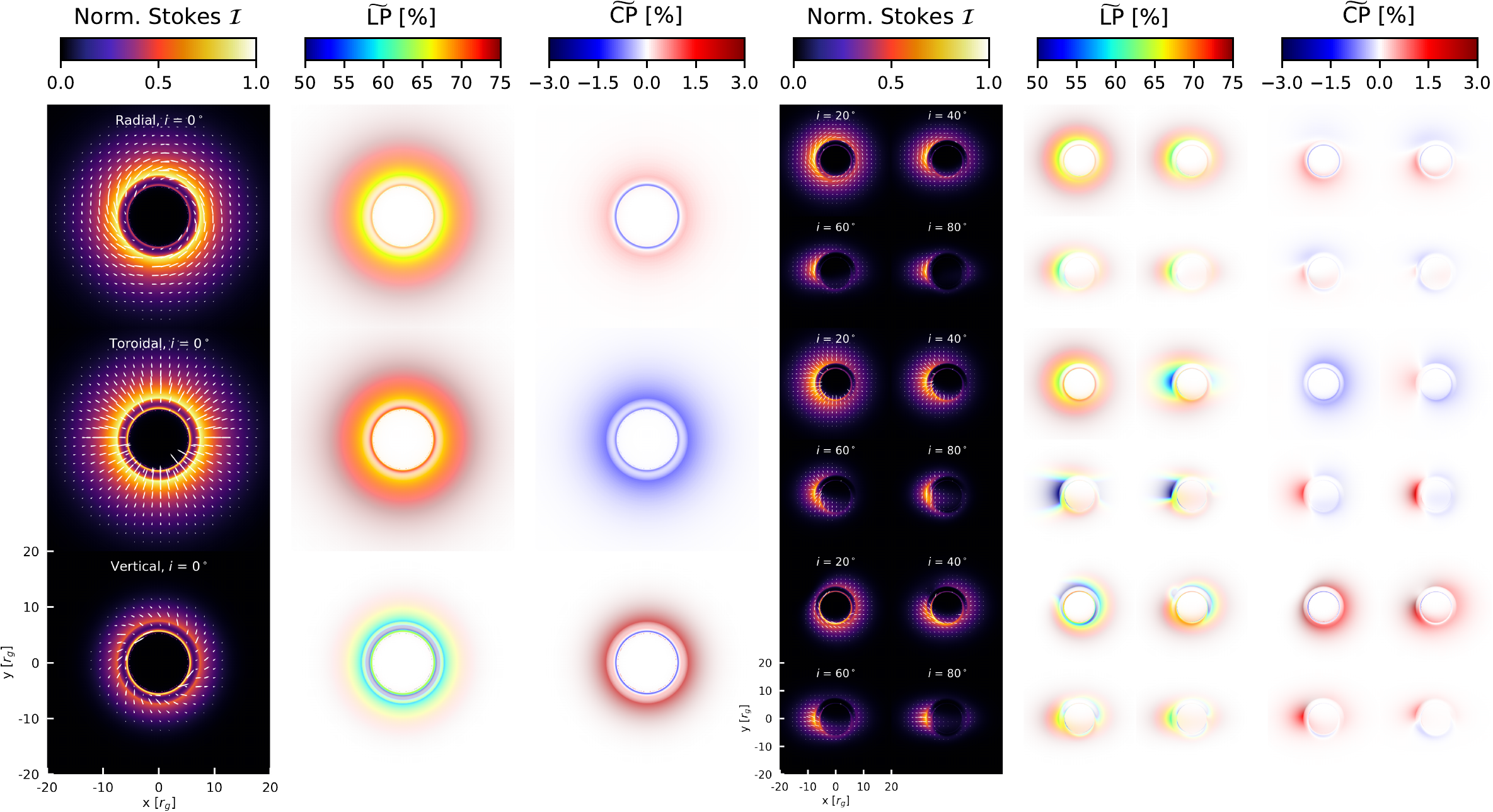}
    \caption{
    Stokes $\mathcal{I}$ and polarimetric images of the background accretion flow as a function of the observer's viewing angle $i$ and magnetic field topology. The \emph{left panel} corresponds to $i=0^\circ$ while $i=20^\circ,40^\circ,60^\circ,80^\circ$ are shown in the \emph{right panel}. Total intensity and polarimetric maps are displayed in the same fashion as in Fig.~\ref{fig:disk_i90_a0}.
    }
    \label{fig:disk_incl_a0}
\end{figure*}

\begin{table}
    \centering
    \caption{Fiducial background accretion flow model parameters}
    \begin{tabular}{cc} \toprule 
    Parameter & Value \\ \midrule 
         $B_0$ & 100 G \\
         $n_0$ & $6 \times 10^6$ cm$^{-3}$\\
         $\Theta_0$ & 200 \\
         $\sigma$ & $0.3$ \\ \bottomrule
    \end{tabular}
    \label{tab:base_model_param}
\end{table}

\begin{table}
    \caption{List of Keplerian orbital periods ($P$) and ISCOs}
    \centering
    \begin{tabular}{ccc} \toprule 
    $a_*$ & P [$r_g/c$] & ISCO [$r_g$] \\ \midrule 
         0.9375 & 235.1 & 2.04\\
         0.5 & 232.4 & 4.23\\
         0 & 229.2 & 6.00\\
         -0.5 & 226.1 & 7.55\\
         -0.9375 & 223.3 & 8.82\\ \bottomrule
    \end{tabular}
    \label{tab:periods}
    \tablefoot{All periods are calculated for a spot orbital radius of $r_\text{spot} = 11$~\rg.}
\end{table}

\section{Results}\label{sec:results}

\subsection{Definitions}
In the following subsections we present results in terms of net fractional linear (LP) and net fractional circular polarization fractions (CP) which are defined as:
\begin{align}
&\text{LP} = \frac{\sqrt{(\sum_i \mathcal{Q})^2+(\sum_i \mathcal{U})^2}}{\sum_i \mathcal{I}}, \\
&\text{CP} = \frac{\sum_i \mathcal{V}}{\sum_i \mathcal{I}}, 
\end{align}
where $\mathcal{I},\mathcal{Q},\mathcal{U},\mathcal{V}$ are the Stokes parameters and the sums ($\sum_i$) are over image pixels. 
We also define:
\begin{align}
&\widetilde{\text{LP}} = \frac{\sqrt{\mathcal{Q}^2+\mathcal{U}^2}}{\mathcal{I}}, \label{eq:LP} \\
&\widetilde{\text{CP}} = \frac{\mathcal{V}}{\mathcal{I}} \label{eq:CP},
\end{align}
where
\lp and \cp denote the linear and circular polarization per pixel.
Additionally, we often discuss the net Stokes parameters' behavior, which are denoted as: 
$F_\nu = \sum_i \mathcal{I}$,
$\text{U} = \sum_i \mathcal{U}$, $\text{Q} = \sum_i \mathcal{Q}$, and $\text{V} = \sum_i \mathcal{V}$.
The net electric vector position angle (EVPA) is defined as:
\begin{equation}
\text{EVPA} = \frac{1}{2} \mathrm{arg} \left( \sum_i \mathcal{Q} + j \sum_i \mathcal{U} \right),
\end{equation}
where $j$ is the imaginary unit and $\text{EVPA} \in (-90^\circ, 90^\circ]$. 
EVPA is measured East of North on the sky.
Therefore, increasing EVPA indicates counter-clockwise rotation of the EVPA. 
$\text{EVPA}=0^\circ$ translates into the polarization tick being aligned with the North-South and $\text{EVPA}=\pm90^\circ$ with East-West direction in the sky (or image). 

In the following subsections, we will outline distinguishing features for the suite of models.
To characterize the spot's position in the image (over time), and thereby the behaviour of $\mathcal{I,Q,U,V}$, we introduce the concept of a location angle $\zeta$.
This angle is a simple aiding quantity to specify the spot's location in the image (and does therefore not outline the geometrical location of the spot in Kerr-Schild coordinates).
For a face-on view of the system ($i=0^\circ$), however, $\zeta$ will be equal to $\phi_\textrm{spot}$, but this is not the case for other inclination angles.
From this viewing angle, the spot's (counterclockwise) orbit will describe a circle. Then, $\zeta=0^\circ$ corresponds to the spot being in the topmost and $\zeta=180^\circ$ to the spot being in the bottom-most positions. 
$\zeta=90^\circ$ and $\zeta=270^\circ$ correspond to the spot being in leftmost and rightmost positions, respectively.

\subsection{Static background emission: Stokes ${\mathcal I}$ and polarization}~\label{res:background}

The left panels of Figure~\ref{fig:disk_i90_a0} display images of the background emission of the zero-spin ($a_*=0$) model for the three different magnetic field configurations as seen from an edge-on viewing angle ($i=90^\circ$).
For this viewing angle, the radial and vertical flux density $F_\nu$ structure is similar -- a bright (relativistically beamed) left side of the disk, a faint (relativistically debeamed) right side of the disk, and a faint lensed ring surrounding the black hole shadow. Notice, that the photon ring around the central black hole shadow is only weakly visible in the total intensity maps. Total intensity images are similar for different magnetic field geometries although the radial and vertical cases look slightly different than the toroidal case.

Linear polarization carries information about the magnetic field topology. Generally, for all magnetic field topologies, \lp is the lowest for the brightest (beamed) region and the highest for the diffuse halo. The toroidal case differs substantially from the others as it has two lower \lp (of $\sim 50\%$) "lobes" above and below the equatorial plane. This behavior is specific to this particular case and is not recovered for other configurations based on inclination or otherwise.
For the radial and vertical cases, the higher flux density regions are less polarized ($60\%$ -- $65\%$).
Polarization ticks in the total intensity maps show the direction of the linear polarization vector $(\mathcal{Q},\mathcal{U})$. These are different for different magnetic field geometries.

For \cp, the structural differences based on magnetic field configuration are even more substantial and they reflect the underlying magnetic field polarity (as we have checked that in our models most of the \cp is produced intrinsically rather then via the Faraday conversion process). The radial case \cp image is predominantly negatively circularly polarized with positive polarization in the Doppler beamed part of the image where the sign (handedness) of the \cp is due to the arbitrarily assumed magnetic field polarity (the inversion of B sign will produce opposite polarity of \cp maps). 
In the toroidal case, the \cp map is positive on the Doppler beamed side and negative elsewhere. 
The vertical case displays a positive \cp for the top half and negative \cp for the lower half resulting in null total CP, as expected. 
Notice that in all cases the magnitude of the fractional circular polarization is much lower when compared to the fractional linear polarization. 

The photon ring (and emission from its direct vicinity) is another feature that differs per magnetic field case. Overall, the photon ring is expected to be relatively less linearly polarized than other regions due to a combinations of plasma (i.e., absorption and Faraday rotation) and gravitational effects, which are further elaborated in \citet{jimenez21}. Although, since photon rings are not the main topic of this paper, we discuss the linear and circular polarization of the photon rings for different magnetic field geometries in  Appendix~\ref{app:photon_ring}.

The right panels of Figure~\ref{fig:disk_i90_a0} display the model background emission for rotating black holes for an edge-on view with $a_*=\pm 0.5,\pm 0.9375$.
There, we find that both the flux density and linear polarization change significantly when the black hole spin becomes negative. The negative spin ($a_* = -0.5,-0.9375$) cases have a higher \lp (and LP) than the positive spin ($a_* = 0.5,0.9375$) cases for all magnetic field configurations. 
This is explained by the increased contribution of relativistic beaming (to the flux) and the higher electron densities, especially in the inner regions, for the positive spin cases.
As we set the electron density to a floor value for radii smaller than the ISCO, the emission from this region will be small. 
Additionally, the emission from these inner regions will experience more Doppler boosting as the (Keplerian) orbital velocity is greater close to the BH.
This explains why the Stokes $\mathcal{I}$ maps of these ($a_* = 0.5, 0.9375$) cases are dominated by a relatively small emission region.
Why the negative spin ($a_* = -0.5,-0.9375$) cases have a bigger emission structure is explained with the same arguments, namely the relatively lower densities and lower orbital velocities at these radii result in an emission structure that is more diffuse and therefore not dominated by a singular feature.
For the fixed magnetic field configuration changing the black hole spin from positive to negative produces significantly different \lp map. Importantly, two observables remain weakly affected by the BH spin value and sign: the linear polarization direction (the EVPA maps) and the \cp maps.

Figure~\ref{fig:disk_incl_a0} shows images of the zero-spin ($a_* = 0$) models as a function of the viewing angle. 
Starting with $i=0^\circ$, the total intensity maps show background disk emission where the inner edge corresponds to the ISCO and the photon ring is located around the black hole shadow. As the viewing angle increases, a characteristic brightness depression appears in the upper left of the flux density maps for the vertical case. 
This "hollow" region is present for viewing angles that are considered, broadly speaking, to be face-on ($ 0^\circ<i \lesssim 45^\circ$) and is due to the Doppler effect which facilities alignment of the wavevector ${\bf k}$ with the magnetic field vector ${\bf B}$ and effectively results in the suppression of the synchrotron emissivity (see Eq. 12 in \citealt{narayan21} or notice that the synchrotron emissivity is $\propto {\bf k} \times {\bf B}$ as measured in the comoving fluid frame). This effect will be critical when describing the variability of the models (presented in Sect.~\ref{res:var_emission_stokesI}).

In Figure~\ref{fig:disk_incl_a0} the fractional linear polarization at $i=0^\circ$ varies per magnetic field configurations. The EVPAs form distinct patterns: for radial fields the polarization ticks are purely azimuthal, for toroidal fields they are purely radial and for vertical field the ticks are nearly azimuthal.
This is consistent with previous predictions \citep{M87PaperVIII}. 
Such EVPA patterns are determined by the relative orientation of the electromagnetic wavevector ${\bf k}$ and magnetic field vector ${\bf B}$ and the effect of relativistic beaming as explained by and in agreement with \citet{narayan21}. For the face-on viewing angle(s) the circular polarization of the radial field configuration is similar to the vertical one, with positively polarized disk emission and photon ring emission with opposite polarization. 
The radial and vertical \cp maps are similar because both of these fields have a poloidal component which is visible at this viewing angle. The toroidal field model, on the other hand, has \cp that is entirely negative. For small viewing angles (up to about $i \leq 30^\circ$) the polarimetric properties of the background emission are similar to the properties of models viewed face-on. 
For $i > 30^\circ$, the linear and circular polarization maps become more complex.
Note that when \cp is produced intrinsically, there is a degeneracy between inverting the viewing angle (from $i=30^\circ$ to $i=150^\circ$, for example) and reversal of the polarity of the field. We extensively comment on this in App. \ref{app:photon_ring}.

We conclude that while the background emission EVPA and \cp depend more on the observer's viewing angle rather than the black hole spin, the background emission \lp is strongly dependent on both the viewing angle and the black hole spin. This gives us a hint as to what polarimetric properties of the bright spot emission may be more geometry (viewing angle and magnetic field) dependent and less dependent on the model details. 

\subsection{Variable emission: Stokes ${\mathcal I}$}~\label{res:var_emission_stokesI}

\begin{figure*}
    \centering
    \includegraphics[width=0.48\textwidth]{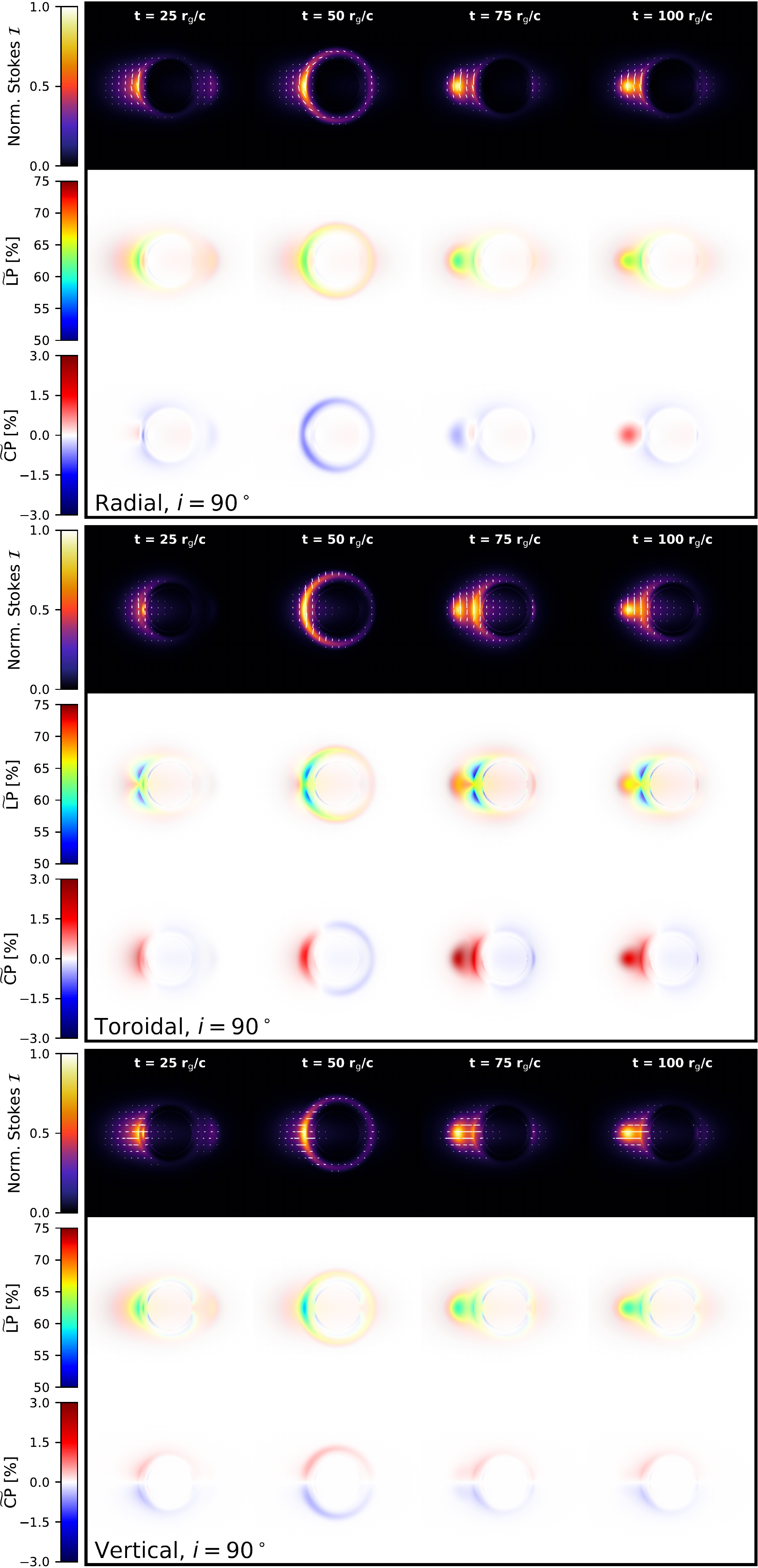}
    \hfill
    \includegraphics[width=0.48\textwidth]{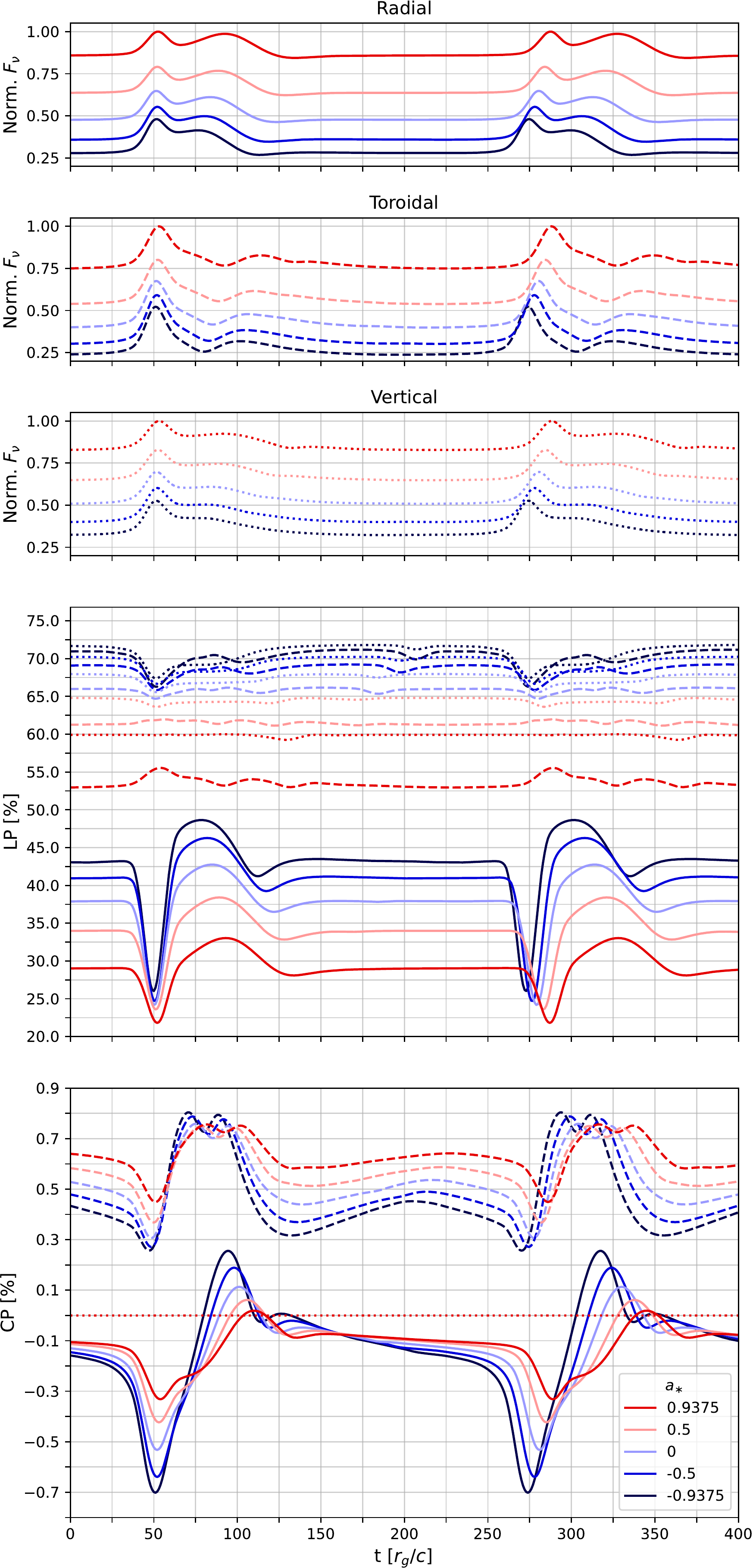}
    \caption{
    {\it Left panels:} An edge-on views ($i = 90^\circ$) of the non-spinning ($a_* = 0$) full (disk + spot) model in radial (\emph{top}), toroidal (\emph{middle}), and vertical (\emph{bottom}) magnetic configurations for a field of view of $40 \, r_g \times 40 \, r_g$. The rows, within each panel, display Stokes $\mathcal{I}$, linear polarization {\fontsize{8pt}{8pt}\selectfont\lpnofont}, and circular polarization {\fontsize{8pt}{8pt}\selectfont\cpnofont} at times $t = 25, 50, 75, 100 \, r_\mathrm{g}/c$. The EVPA orientation is represented with white ticks on top of the Stokes $\mathcal{I}$ panels.
    Their length is determined by the {\fontsize{8pt}{8pt}\selectfont\lpnofont} strength. 
    {\it Right panels:} Comparison of the evolution of the flux density (top), the net linear polarization (middle) and the net circular polarization (bottom) over time for a radial (solid), toroidal (dashed), and vertical (dotted) magnetic field configuration shown in the left panels. Here, we additionally plot light curves for black holes with $a_*\neq0$ (different black hole spins are color coded).}
    \label{fig:spot_i90_a0}
\end{figure*}

\begin{figure*}
    \centering
    \includegraphics[width=0.48\textwidth]{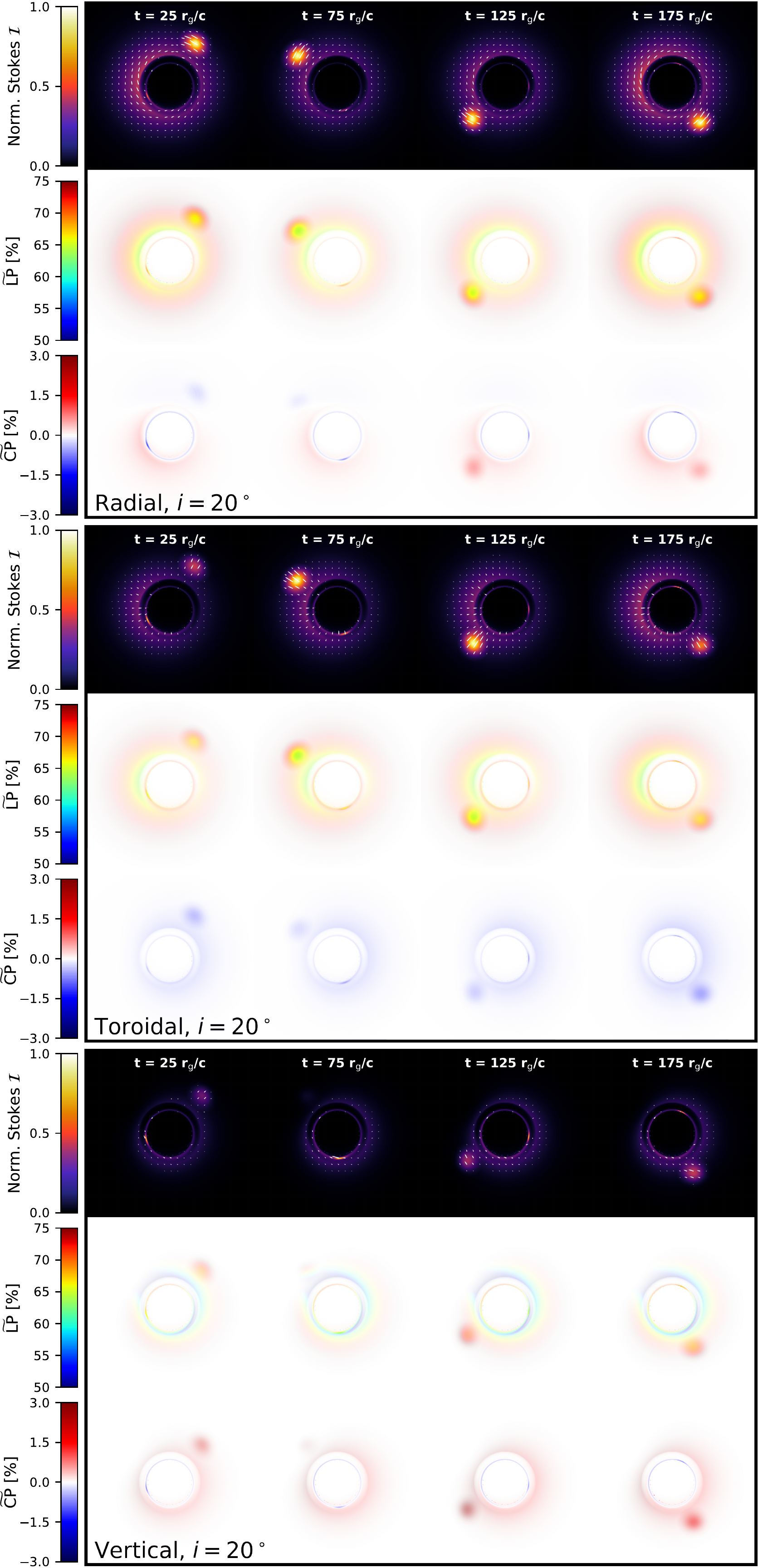}
    \hfill
    \includegraphics[width=0.48\textwidth]{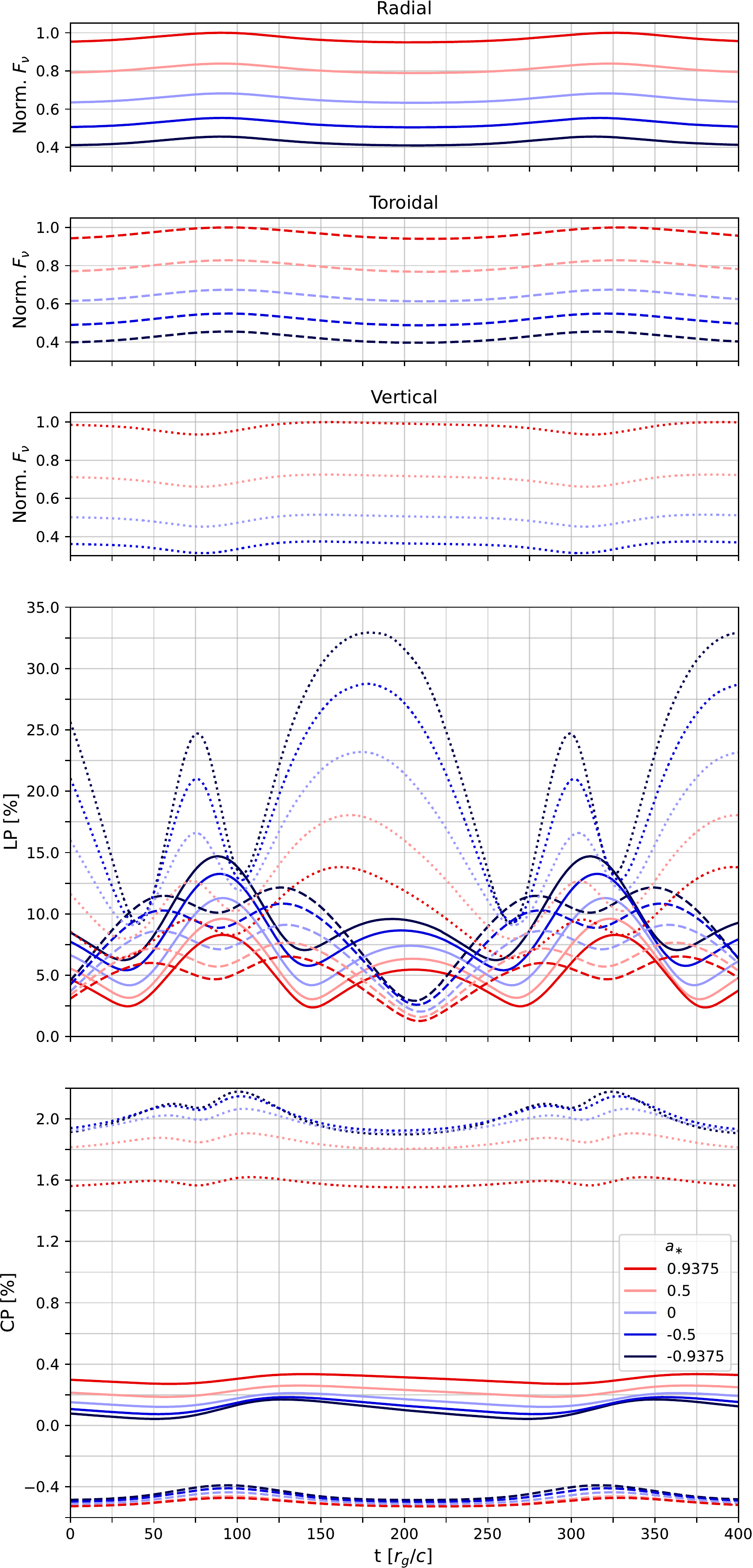}
    \caption{The same description as in Fig.~\ref{fig:spot_i90_a0} except for the viewing angle which is $i=20^\circ$ and we show maps corresponding to times $t = 25, 75, 125, 175 \, r_\mathrm{g}/c$.}
    \label{fig:spot_i20_a0}
\end{figure*}

Before we discuss the polarimetric findings for an orbiting spot, we will outline a number of characteristic features that appear in Stokes $\mathcal{I}$. We discuss in more detail two cases with an edge-on viewing angle and a nearly face-on viewing angle.  

In Fig.~\ref{fig:spot_i90_a0}, in left panels, we show model images with $a_*=0$ and all three magnetic field configurations for an edge-on viewing angle ($i=90^\circ$). These images are naturally paired with the light curves shown in the right panels in Fig.~\ref{fig:spot_i90_a0}. Our full (background plus spot) model describes a periodic system, hence the light curves can be arbitrarily shifted.
Therefore, we align the light curves in such a way that the spot's position is similar between different magnetic field cases. 
The maximum flux density, in the light curve shown in Fig.~\ref{fig:spot_i90_a0}, always corresponds to the formation of an "Einstein ring"-like image at edge-on viewing angles.
Then, as the Einstein ring disappears, we perceive a decrease in flux density.
This decrease is slight for the radial case, more substantial for the vertical case, and significant for the toroidal case.
After this decrease, the flux density will increase again as the spot moves in view and is relativistically beamed (or Doppler boosted) towards the observer.
The recovery of the flux density is also highly dependent on the magnetic field geometry. 
The second peak of the radial case almost reaches the level of the first maximum.
For the vertical case, we see that the second maxima recovers to about half of the first maximum amplitude.
The toroidal case only increases to a third of the first maximum. These `recoveries' are all quantified with respect to the corresponding curves' minimum.

Figure~\ref{fig:spot_i20_a0} shows the same models as in Fig.~\ref{fig:spot_i90_a0} but for the nearly face-on viewing angle ($i=20^\circ$). Here the observer has an unobstructed view of the spot during the entire orbit.
As the emission of the spot is not heavily distorted by strong gravitational lensing, we recover a simpler, single-peaked, structure in the light curve.
The flux density maximum occurs when the spot's emission is beamed towards the observer (at $t\approx 100 \, r_g/c$ or $\zeta\approx45^\circ$).
However, contrary to the edge-on case, the vertical ($i=20^\circ$) case has a {\it minimum} at this point in the spot's orbit, where the wave vector ($\bm{k}$) and the magnetic field vector ($\bm{B}$) are parallel to each other and the synchrotron emissivity emission becomes negligible.

How sensitive is the variable component of the total intensity to the black hole spin?
For the edge-on cases (in Fig. \ref{fig:spot_i90_a0}), we find that the contribution of the spot is largest for the $a_*=-0.9375$ case, with (up to) $50\%$ being determined by the variable component, and smallest for the $a_*=0.9375$ case, with only (up to) $25\%$ being determined by the variable component.
As the orbital velocity increases as $a_*$ decreases, the contribution of the spot increases as a result of relativistic beaming.
Additionally, this difference in contribution to the total flux density is also partly explained by the $r<r_\textrm{ISCO}$ cut we apply to the disk background.
For the nearly face-on cases (in Fig. \ref{fig:spot_i20_a0}), the spot's contribution to the total flux density is less dependent on spin and seems to be roughly $10\% - 20\%$.
As the observer inclination no longer lines up with spot's orbital plane, relativistic beaming will be less strong than for the edge-on case. It also no longer has the first (greater) flux density peak which corresponds to the occurrence of the Einstein ring.

\subsection{Variable emission: polarization}\label{res:variable_emission_polarization}

\subsubsection{Polarimetric signatures of bright spots: \quloops}\label{res:QUloops}

\begin{figure*}
    \centering
    \includegraphics[width=\textwidth]{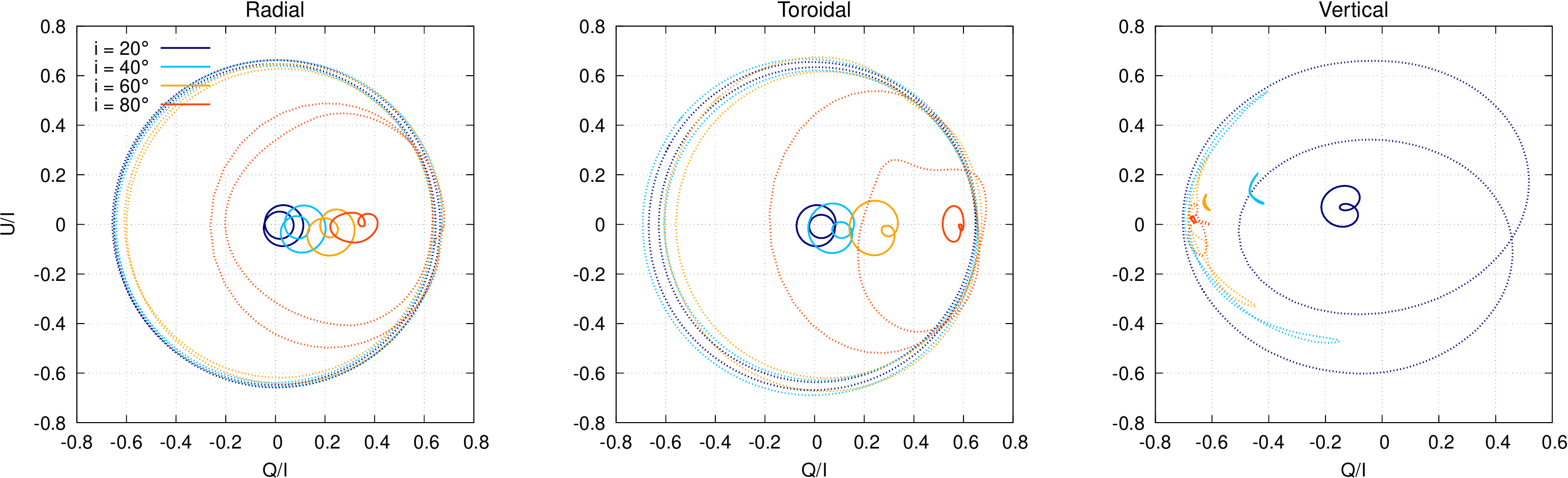}
    \caption{Normalized $\mathcal{Q}$-$\mathcal{U}$ diagrams for the three basic magnetic field configurations and a black hole spin of $a_*=0$. Both the spot-only (dotted lines) and full models' (solid lines) \quloops are shown for four viewing angles $i=20^\circ, 40^\circ, 60^\circ, 80^\circ$.}
    \label{fig:QU_spot_disk}
\end{figure*}

The evolution of EVPA is best described using \qu diagrams.
Looping structure in ${\mathcal U}$ versus ${\mathcal Q}$ diagram (or put simply as "\qu diagram") is a hallmark signature of an orbiting hot spot 
\citep{marrone06phd}. 
However, one could argue that the (disk) background emission is equally important in shaping the loop structure and position in the \qu diagram. This is illustrated in Fig.~\ref{fig:QU_spot_disk}, where we show \qu diagrams for both the spot-only and the full (disk+spot) configurations for \emph{radial} (left panel), \emph{toroidal} (middle panel), and \emph{vertical} (right panel) magnetic fields. 
Without commenting yet on the loops' finer substructure, we find that the spot-only loops are greater in size and broadly describe a circle for the radial and toroidal configurations which are largely independent of viewing angle.
The vertical configuration is more dependent on the system's inclination angle - only the $i=20^\circ$ case shows looping structure. 
The smaller size of the full configuration's \quloops is largely due to the compensation with the flux density $F_\nu$.
As was commented in Sect. \ref{res:background}, the spot-only configuration has much smaller flux densities than the full model. 
Surprisingly, there are clear structural differences between the spot-only and full model descriptions for the ${\mathcal Q}/{\mathcal I}$,${\mathcal U}/{\mathcal I}$ loops, whereas this is not the case for the \quloops (unnormalized diagrams where ${\mathcal Q}$ and ${\mathcal U}$ are shown in units of Jansky).
For the latter diagrams, which are currently not shown, we find an identical \qu structure for both the spot-only and full configurations but they are displaced from each other by an offset as a result of the disk background.
Therefore, the clear differences in the ${\mathcal Q}/{\mathcal I}$,${\mathcal U}/{\mathcal I}$ diagrams (between the spot-only and full models) originate from the disk component that dominates the flux density (or Stokes $\mathcal{I}$), especially when the spot is receding from the observer.
The (near) circular structure in the ${\mathcal Q}/{\mathcal I}$,${\mathcal U}/{\mathcal I}$ for the spot-only cases is an indication that the variation in ${\mathcal I}$, ${\mathcal Q}$, and ${\mathcal U}$ are largely (but not perfectly) correlated. This circular symmetry for the radial and toroidal cases is, mainly, broken by relativistic beaming and, partially, by Faraday rotation -- see Appendix \ref{app:lin_cir_polarization}.

\begin{figure*}
    \centering
    \includegraphics[width=0.238\textheight]{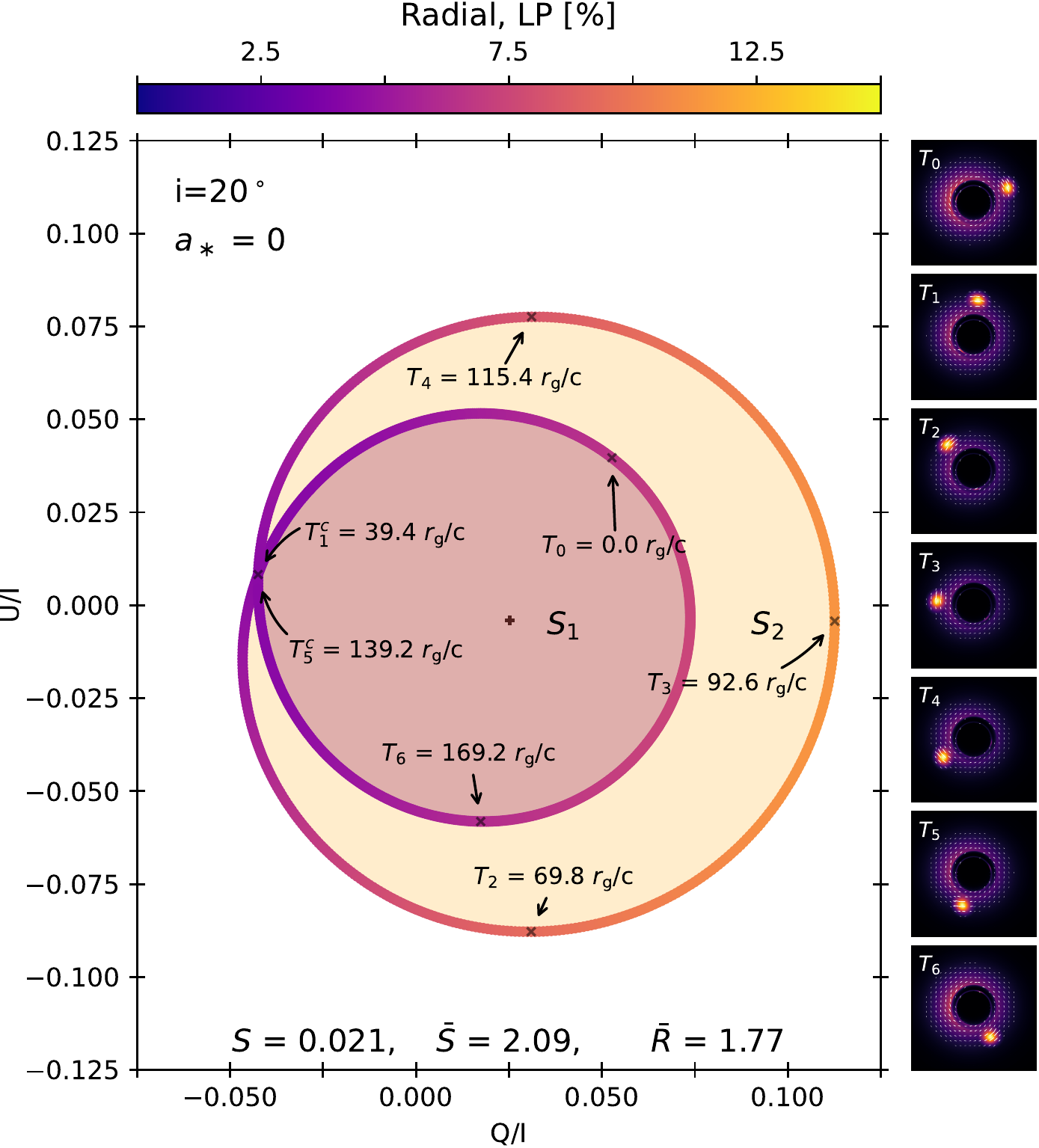}    
    \hfill
    \includegraphics[width=0.238\textheight]{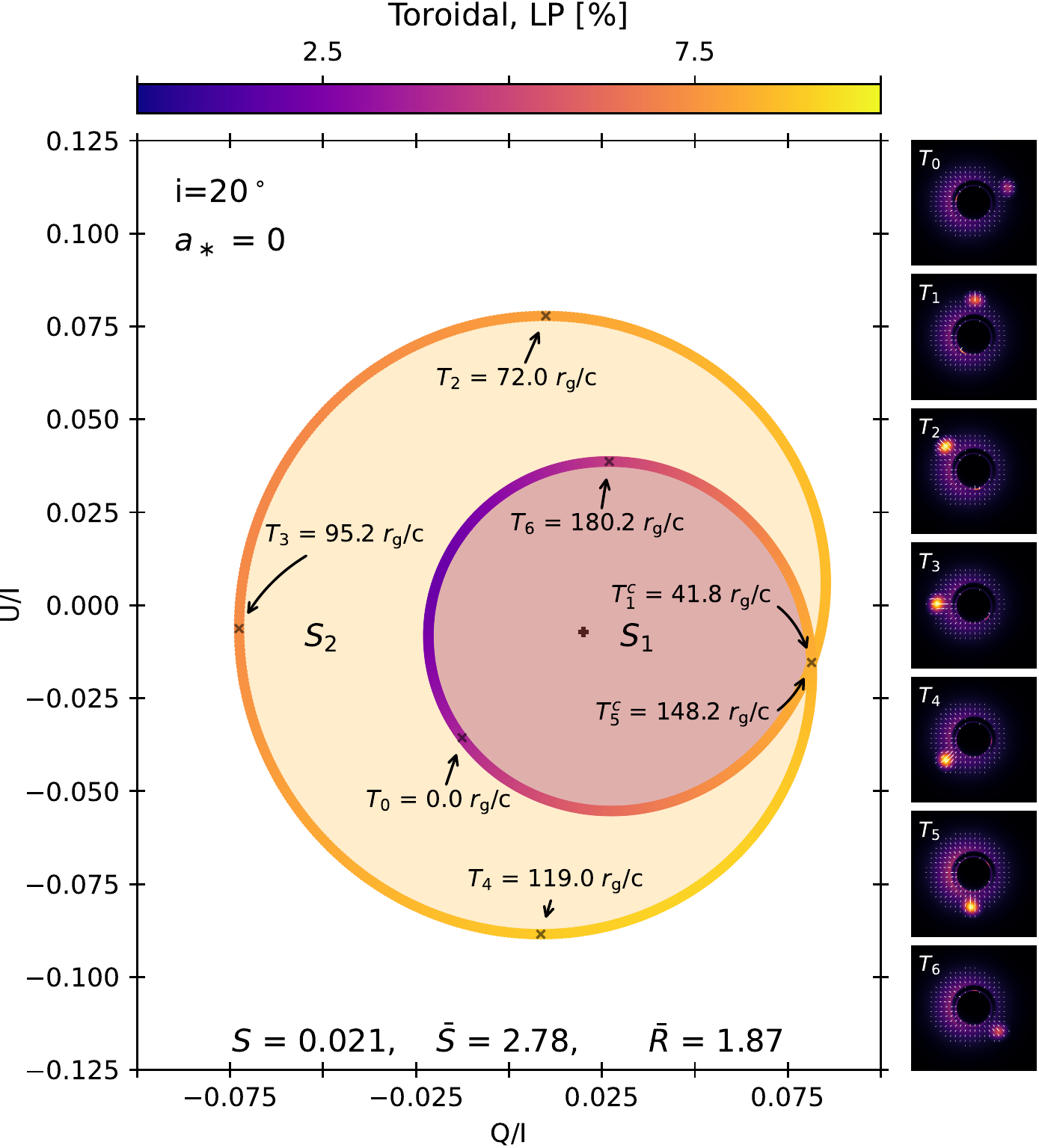}
    \hfill
    \includegraphics[width=0.225\textheight]{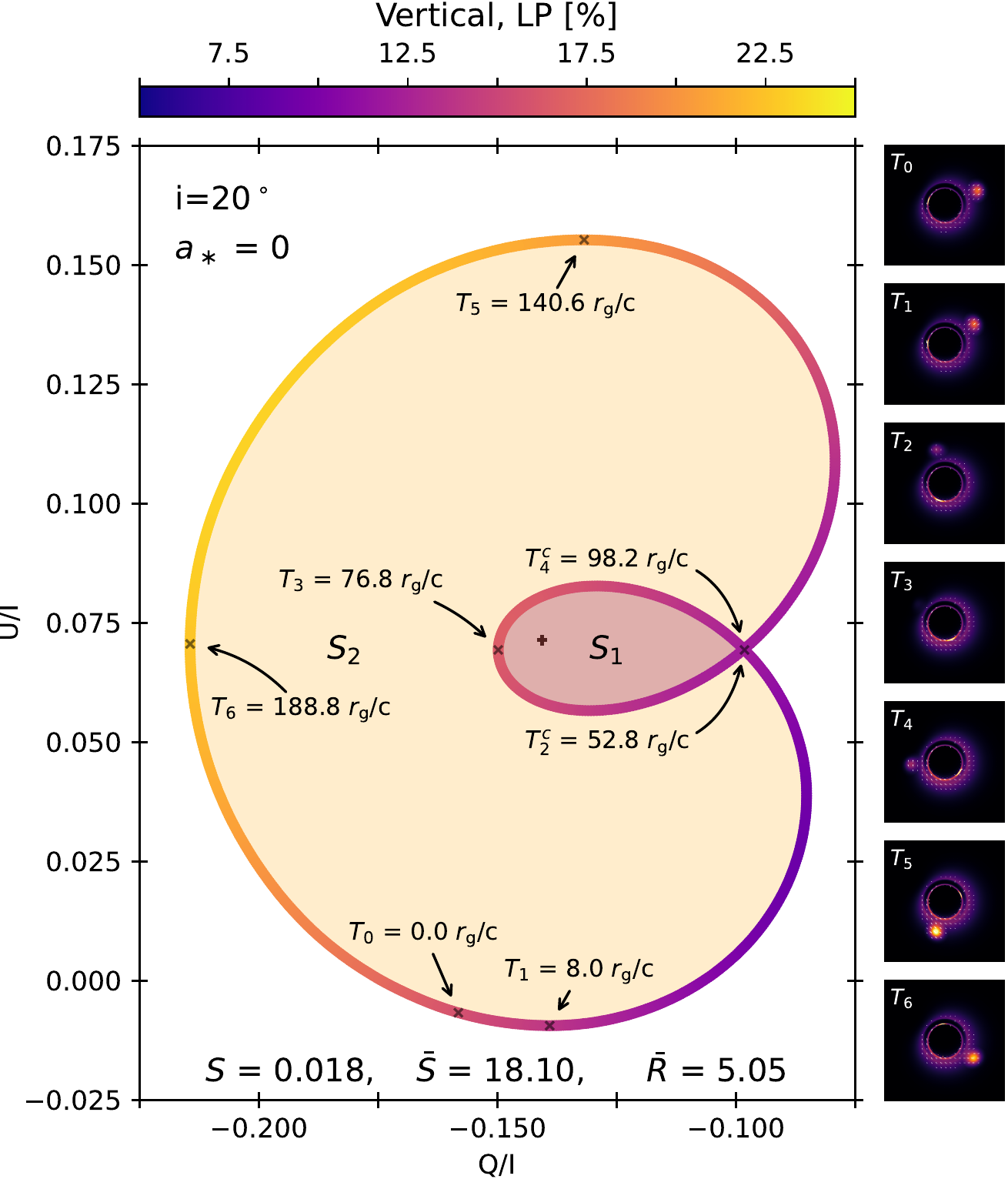}
    \caption{Schematic diagrams of normalized \quloops of the full models (for $i=20^\circ$, $a_*=0$) with the radial, toroidal, and vertical magnetic field configurations in the \emph{left}, \emph{middle}, and \emph{right} panel, respectively. The loop color encodes the model LP. The inner loop surface $S_1$ (in \emph{pink}) and rest of surface (excluding $S_1$) enclosed by the entire \quloop structure is denoted by $S_2$ (in \emph{yellow}). The total surface is $S = S_1 + S_2$, surface fraction is $\bar{S} = S/S_1$ and time spent in inner loop with respect to the total period is $\bar{R} = P / (T_4^c-T_2^c)$ \citep[similar to][]{wielgus22polar}. The `+' symbol marks the normalized \qu value of the static background emission. All shown loops are traversed in the counterclockwise direction.}
    \label{fig:QUschematic}
\end{figure*}

The properties of the \quloops depend on a number of model parameters.
Although, before we describe how the \quloops depend on them, we will introduce metrics to quantitatively describe the loops.
Figure~\ref{fig:QUschematic} displays a schematic overview of the typical \quloop structure that appears in our models. 
Similar to \citet{wielgus22polar}, we describe the loops (i.e., a pretzel-like shape) using two dimensionless metrics. 
Another, additional, metric we consider (which \citet{wielgus22polar} does not) is the total area of the loop structure in the {\bf \qu} space, $S=S_1 + S_2$. 
The first of the dimensionless metrics is the ratio of total loop area ($S$) to small loop area ($S_1$), which is denoted as $\bar{S} = S/S_1$. 
The last metric is the inner loop timing ratio $\bar{R} = P / (T^c_\text{exit} - T^c_\text{enter})$.
Here, the $T^c_\text{enter}$ is the time at which the {\bf \qu} signature enters the inner loop and $T^c_\text{exit}$ is the time at which it exits the inner loop. 
Then, $\bar{R}$ gives the period {\bf $P$} divided by the time spent within the inner loop.

By inspecting Fig.~\ref{fig:QUschematic} one finds that for radial and toroidal magnetic fields the \qu signature will be following the inner loop when the spot is receding from the observer ($\zeta = 180^\circ$ to $\zeta = 360^\circ = 0^\circ$). The \qu signature will follow the outer loop as the spot approaches the observer  ($\zeta = 0^\circ$ to $\zeta = 180^\circ$).
For the vertical magnetic field case, this behaviour is quite the opposite. Roughly speaking, the vertical magnetic field's \qu signature will traverse the left half of the \quloop, roughly the outer loop, for $\zeta \approx 160^\circ$ to $320^\circ$.
There, generally speaking, the spot is receding with respect to the observer.
Therefore, the right half of the \qu diagram, including the inner loop, corresponds to the approaching spot. 
The \qu signature follows the inner loop when the contribution of the spot (almost completely) disappears in the region where synchrotron emission is suppressed (as at $T_3$ we only can see a lensed image of the spot). 
The vertical cases that do not possess an inner \quloop (for $i\gtrsim25^\circ$) are still affected by the emission suppression region.
Then, the indentation (where the inner loop would have been) corresponds to the part of the spot's orbit where the emission is suppressed (for $i=25^\circ$).
When we look at Fig. \ref{fig:QUschematic}, we find that for the vertical cases only a relatively small part of the spot's orbital period is spent inside the inner \quloop when compared to toroidal/radial cases. 

Next, we investigate the \quloop characteristics, (using the metrics defined above) as a function of a few model parameters such as magnetic field geometry, observers' viewing angle, radius of the spot, black hole spin, and plasma parameters (we consider only small deviations from our fiducial values of density, temperature and magnetic field strength). We summarize our main findings in the paragraphs below.

\begin{figure}
    \centering
    \includegraphics[width=0.48\textwidth]{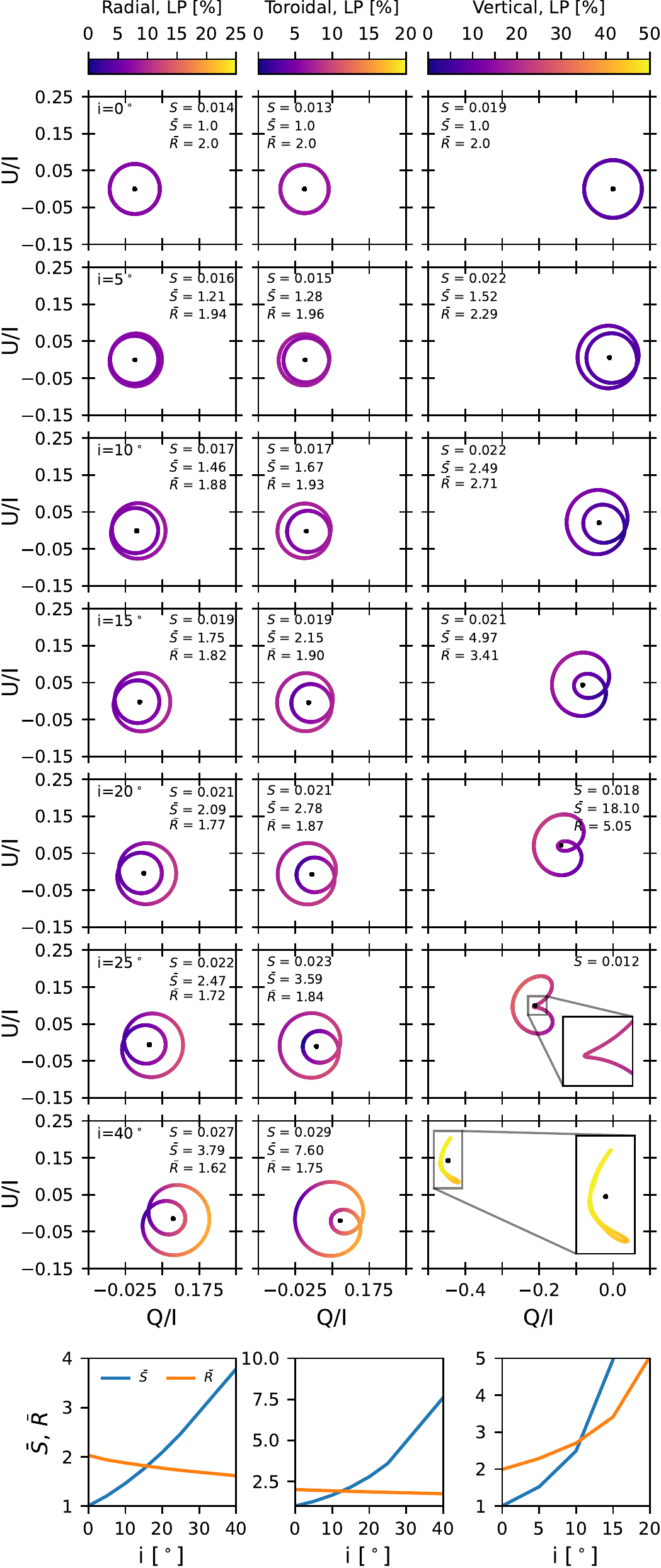}   
    \caption{Normalized \qu diagrams of the full model colored according to their LP fraction as a function of increasing inclination $i$ (panels from top to bottom).
    Each column corresponds to a different magnetic field configuration. All models have $a_*=0$. While for vertical magnetic field the \quloops disappear for $i \gtrsim 25^\circ$, for other field geometries 
    they are present for all viewing angles except exact edge-on cases (see also Figure~\ref{fig:QU_spot_disk}).
    The bottom row shows the behavior of $\bar{S}$ (in blue) and $\bar{R}$ (in orange) as a function of inclination $i$. The observables $S$, $\bar{S}$, and $\bar{R}$ are explained in Fig.~\ref{fig:QUschematic}. 
    }  
    \label{fig:QULP_incl}
\end{figure}

\paragraph{Viewing angle:}
Figure~\ref{fig:QULP_incl} shows that the dominant 
decider for \quloops shape is the inclination angle of the system. For a face-on viewing angle ($i=0^\circ$), we will only recover circle-shaped loops, regardless of the field configuration. Notice that for a single orbital period this circle is traversed two times, which indicates that the inner and outer loops are still present but have the same size (corresponding to $\bar{S}=1$ and $\bar{R}=2$). Notice also that all our loops are traversed in the counter-clockwise direction because the spot is orbiting counter-clockwise on the sky (we only show results for viewing angles $i>90^\circ$).
As the inclination $i$ increases, the properties of the \quloops strongly depend on the underlying magnetic field geometry.
For the radial and toroidal cases, for increasing $i$, the inner loop starts to shrink while the outer loop remains the same size. For the vertical cases, we find that the inner loop disappears already for $i \gtrsim 25^\circ$ (with the last inner loop structure being present at $i \approx 24^\circ$ to be more exact). For radial and toroidal cases the loop persist for all viewing angles except the exact edge-on case. 
For all three magnetic field topologies, $\bar{S}$ increases with increasing viewing angle, but for vertical field the dependency is stronger. 
$\bar{R}$ can be used to discriminate between radial/toroidal and vertical cases: $\bar{R}$ slowly decreases with $i$ for radial/toroidal fields, while $\bar{R}$ increases with $i$ for vertical field. The sharp increase of $\bar{R}$ as a function of $i$ is related to the duration of the ${\bf k} \times {\bf B}$ suppression as a result of relativistic beaming which is a strong function of viewing angle. 

\begin{figure}
    \centering
    \includegraphics[height=0.93\textheight]{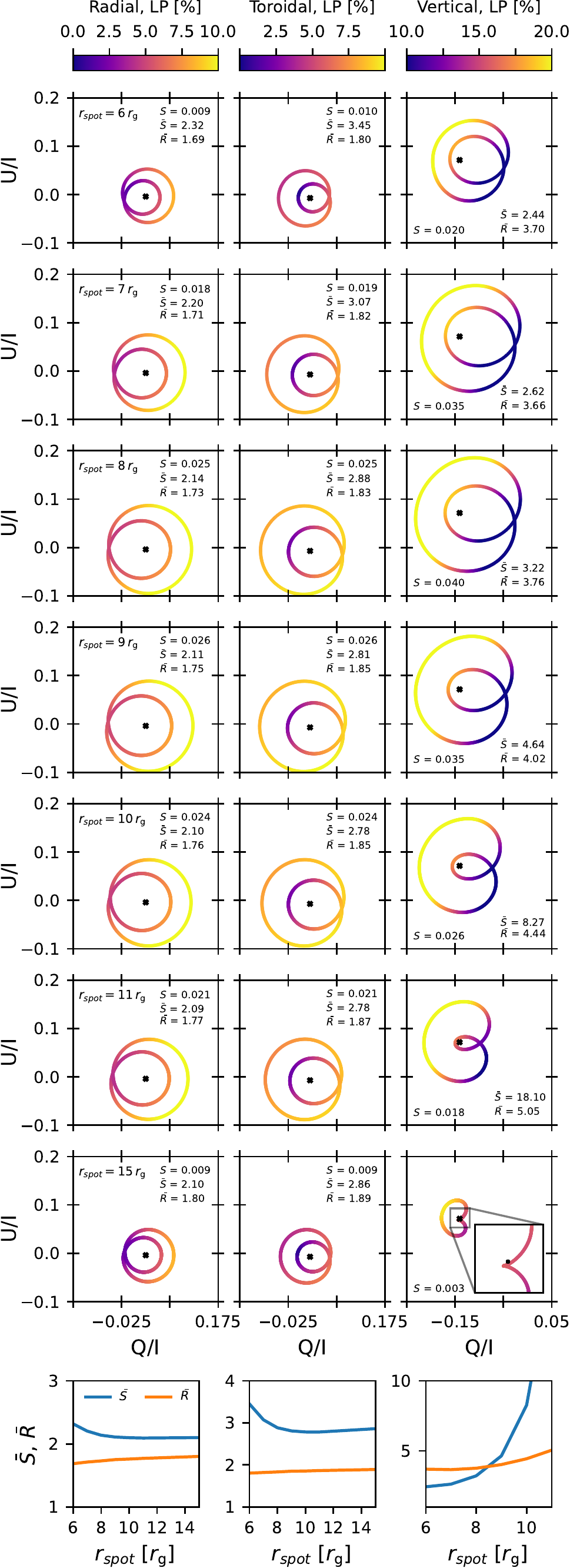}
    \caption{Normalized \qu diagrams colored according to their LP fraction. Each row corresponds to a different orbital radius $r_{spot}$. The inclination and black hole spin are set to $i=20^\circ$ and $a_*=0$, respectively. The bottom row shows the behavior of $\bar{S}$ (in blue) and $\bar{R}$ (in orange) as a function of orbital radius $r_\text{spot}$. The observables $S$, $\bar{S}$, and $\bar{R}$ are explained in Fig.~\ref{fig:QUschematic}.
    }
    \label{fig:QULP_rspot}
\end{figure}

\paragraph{Spot orbital radius:}
Figure~\ref{fig:QULP_rspot} displays the \quloop structure as a function of the spot's orbiting radius $r_\text{spot}$. The viewing angle and spin are fixed to $20^\circ$ and $a_*=0$, respectively.
For the vertical case with $r_\text{spot} \leqslant 11$ \rg, we still find an inner loop, but the inner loop decreases in size for increasing $r_\text{spot}$ until it is no longer present which becomes clear for $r_\text{spot} = 15$ \rg. 
For the radial and toroidal cases, we recover the double \quloop structure for all $r_\text{spot}$ values.
Their total surface $S$ increases till $r_\text{spot} = 9$ \rg after which it decreases. 
$\bar{R}$ seems insensitive to different $r_\text{spot}$ values, while $\bar{S}$ is almost constant except for $r_\text{spot}=6$ \rg.
This slight increase in $\bar{S}$ is likely related to cut-off at ISCO radius.
The vertical magnetic field cases display a largely constant $\bar{R}$ for increasing $r_\text{spot}$, but $\bar{S}$ does increase with $r_\text{spot}$.
This behavior has the same explanation as was stated in the previous paragraph, i.e. the orbital speed increases for decreasing $r_\text{spot}$.

\begin{figure}
    \centering
    \includegraphics[width=0.48\textwidth]{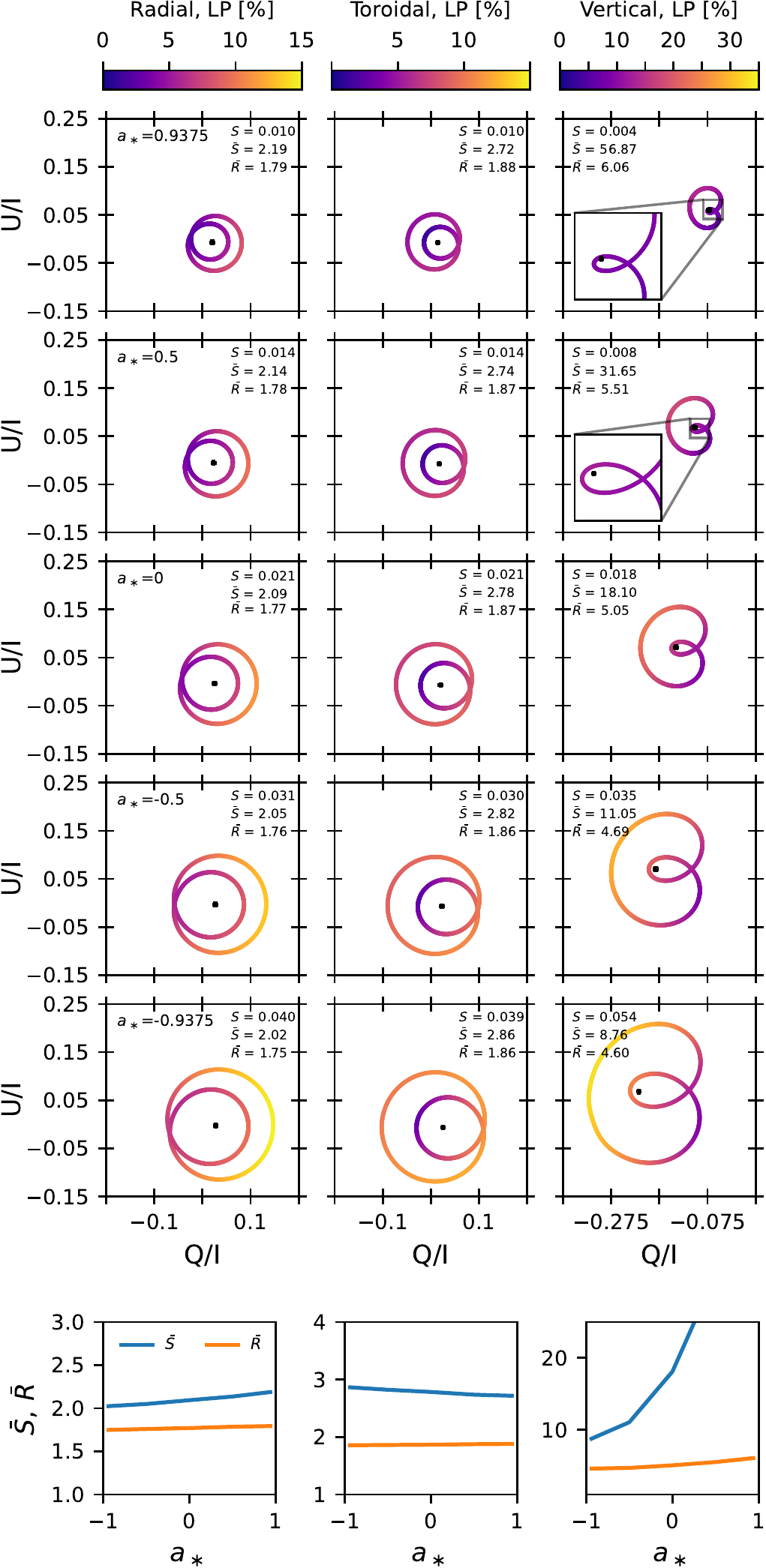}
    \caption{Normalized \qu diagrams colored according to their LP fraction. The inclination is set to $i=20^\circ$. Each row corresponds to a different black hole spin parameter $a_*$. The bottom row shows the behavior of $\bar{S}$ (in blue) and $\bar{R}$ (in orange) as a function of $a_*$. The observables $S$, $\bar{S}$, and $\bar{R}$ are explained in Fig.~\ref{fig:QUschematic}.
    }
    \label{fig:QULP_i20}
\end{figure}

\begin{figure}
    \centering
    \includegraphics[width=0.48\textwidth]{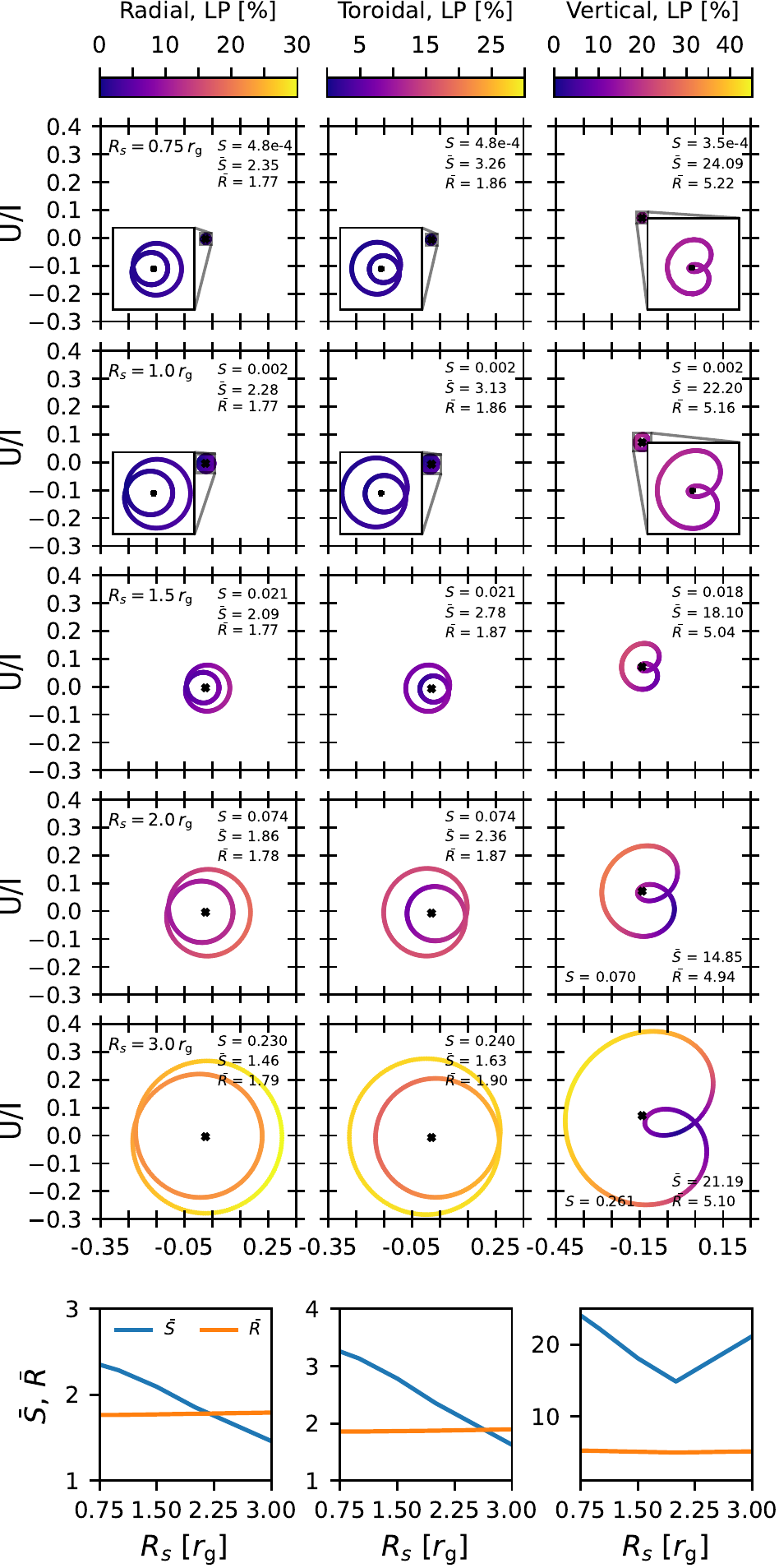}
    \caption{Normalized \qu diagrams colored according to their LP fraction for $i=20^\circ$ and $a_*=0$ cases. Each row corresponds to different spot sizes ($R_s$). The default in the rest of this work is $R_s = 1.5$ \rg. The observables $S$, $\bar{S}$, and $\bar{R}$ are explained in Fig.~\ref{fig:QUschematic}.
    }
    \label{fig:QULP_spotsize_i20}
\end{figure}

\begin{figure}
    \centering
    \includegraphics[width=0.48\textwidth]{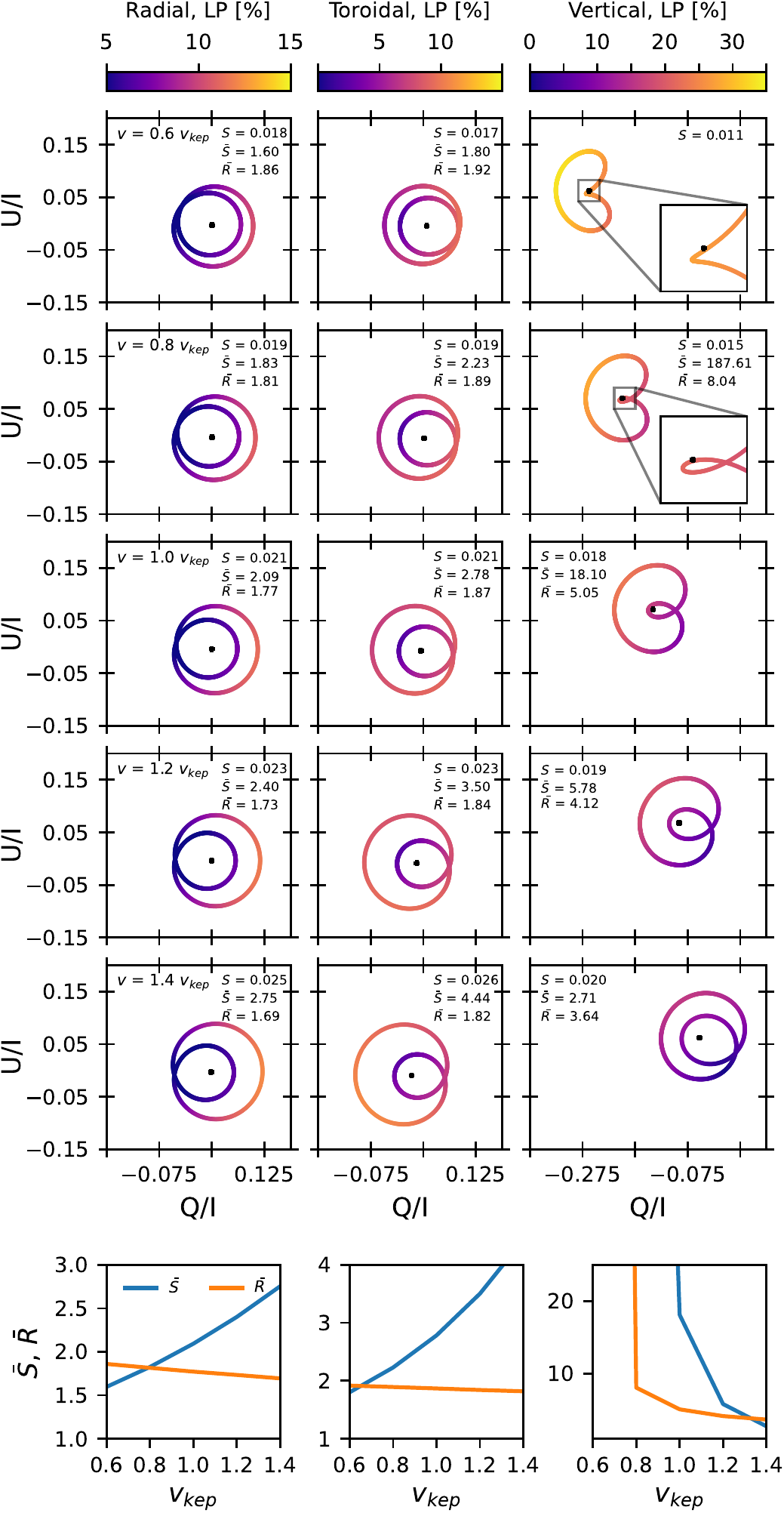}
    \caption{Normalized \qu diagrams colored according to their LP fraction for $i=20^\circ$ and $a_*=0$ cases. Each row corresponds to different orbital velocities denoted as a fraction of the standard velocity in this work which is Keplerian ($v_{kep}$). The observables $S$, $\bar{S}$, and $\bar{R}$ are explained in Fig. \ref{fig:QUschematic}.
    }
    \label{fig:QULP_keplerian_i20}
\end{figure}

\begin{figure}
    \centering
    \includegraphics[width=0.48\textwidth]{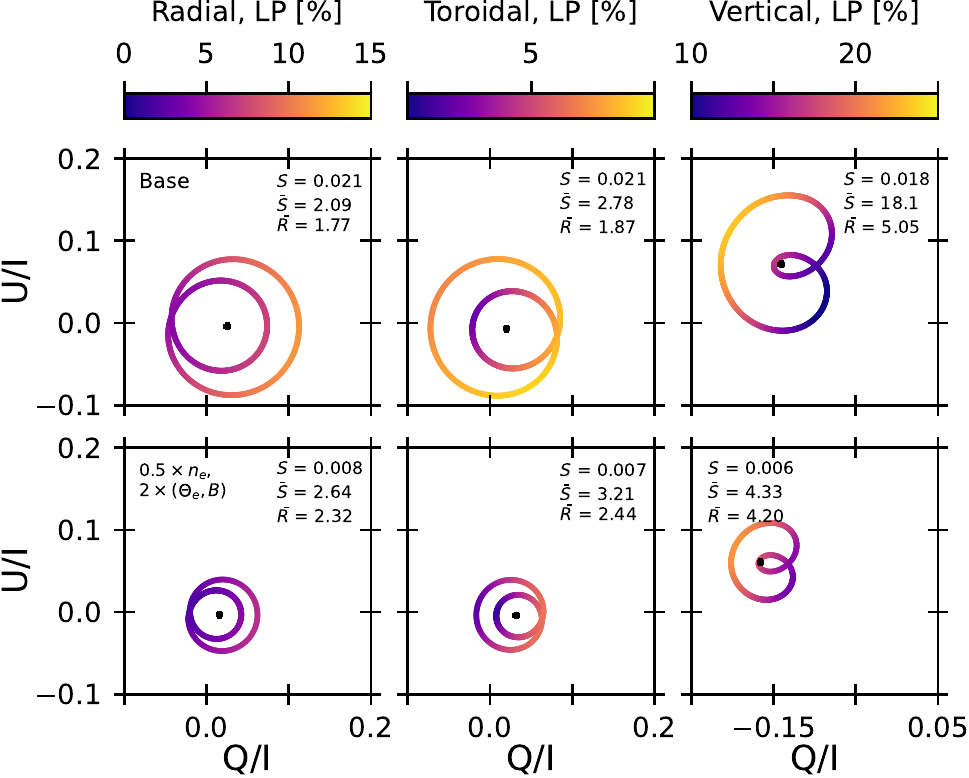}
    \caption{Normalized \qu diagrams colored according to their LP fraction. The \emph{top row} corresponds to the base model with $r_\text{spot} = 11$ \rg, $a_*=0$, and $i=20^\circ$ and the \emph{bottow row} shows some exploratory models with doubling of $\Theta_e$ and $B$ and halved $n_e$. The observables $S$, $\bar{S}$, and $\bar{R}$ are explained in Fig. \ref{fig:QUschematic}.
    }
    \label{fig:QULP_other_i20}
\end{figure}

\paragraph{Black hole spin:}
Figure~\ref{fig:QULP_i20} displays \quloops and their properties as a function of black hole spin.
We find that the size of the \quloop is the greatest for the lowest spin ($a_*=-0.9375$).
This behavior is a direct result of our model assumptions.
We assume a low floor electron density $n_e$ at radii below the ISCO.
The contribution of the disk background will therefore vary between spin-cases, while the contribution of spot will remain unaffected.  
Hence, the $a_* = 0.9375$ case has the largest disk background contribution, whereas the $a_* = -0.9375$ case has the lowest disk background contribution.
This is an explanation for the increase in \qu loop size ($S$) with decreasing spin parameter. 
The toroidal cases' \qu loops are similar between the five spin cases.
To a lesser extent this can also be stated about the vertical cases.
However, for the most positive ($a_*=0.9375$) spin-case, we observe that the upper part of the loop is relatively larger with respect to the lower part. For the most negative ($a_*=-0.9375$) spin, this is the other way around but less pronounced. The other spin-cases ($a_* = 0, \pm0.5$) display a clearer symmetry between upper and lower half of the loop.

We find that the total surface $S$ increases when the spin decreases for all cases.
For the radial and toroidal cases, we recover an almost constant $\bar{S}$ and $\bar{R}$ similar to what we saw in the previous paragraph (for $r_\text{spot}$).
For the vertical case, we find that $\bar{R}$ is only a weak function of spin, but $\bar{S}$ is a exceptionally strong function of spin.
Given the immutable nature of the \quloop geometry and the strong dependence of $\bar{S}$ (and $S$) on spin, it would be possible to assess the black hole spin $a_*$ of the system for the vertical case (if the inclination is well-constrained). 
For the radial and toroidal cases, this is less promising. 
However, the total surface $S$ displays a stronger dependence on spin than it has shown for the other investigated quantities.
We do recover near identical $S$ between the radial and toroidal fields (as we have seen previously for other \qu figures). 

Lastly, we discuss the radial cases in the left column of Fig.~\ref{fig:QULP_i20} in more detail.
Each \quloop has a point where one crosses from the inner to the outer loop. This point will henceforth be referred to as the crossing-point.
For the radial cases, we find that the crossing-point lies in the upper half of the \quloop for the positive spin-cases.
However, as the spin decreases, the crossing-point gradually moves towards the lower half of the \quloop. 
This behavior is still present if we remove the background, which indicates that it is likely related to the underlying spacetime.

\paragraph{Spot size:}
Figure \ref{fig:QULP_spotsize_i20} displays the dependency of \quloops on the spot size parameter $R_s$.
Variations in $R_s$ bring about significant changes in, especially, \quloop size but not so much the structure, except for the $R_s = 3.0$ \rg.
Note that the ($\mathcal{Q}/\mathcal{I},\mathcal{U}/\mathcal{I}$) domain shown in this figure is significantly greater than for the other \qu figures.
The spot's contribution to the overall flux density increases with increasing spot size.
This is reflected in the total loop size $S$.
$\bar{R}$ remains independent of the spot size.
$\bar{S}$, on the other hand, tends to decrease for increasing $R_s$ in the radial and toroidal magnetic field configurations.
As was established in the previous paragraphs (and shown schematically in Fig. \ref{fig:QUschematic}), the inner \quloop is followed when the spot is receding from the observer.
For the vertical, the inner loop has a different origin, namely a region where $\bm{k} \times \bm{B} \sim 0$.
This explains why the $R_s = 3.0$ \rg case breaks the trend of steadily decreasing $\bar{S}$ for increasing $R_s$ as the spot has become sufficiently large to no longer be fully suppressed in the $\bm{k} \times \bm{B} \sim 0$ region.
Note that the relative growth of the inner loop size indicates that the spot's emission starts eclipsing the contribution of the disk. If this were to continue, then we find that we end up with \quloop behavior that is similar to what we have seen in Fig. \ref{fig:QU_spot_disk} for the spot-only case(s). 

\paragraph{Orbital velocity:}
Figure \ref{fig:QULP_keplerian_i20} displays the \quloop behavior for deviations from the base model's Keplerian orbital velocity (shown in the middle row).
We modify the orbital velocity of the entire system (disk + spot) with a factor that ranges from 0.6 (sub-Keplerian) to 1.4 (super-Keplerian).
For the radial and toroidal cases, where the inner and outer \quloop correspond to the receding and approaching spot, we find that the inner and outer loop become more similar in size (lower $\bar{S}$ but higher $\bar{R}$) for decreasing orbital speed. 
This behavior is explained by the increased (decreased) contribution of relativistic beaming for increasing (decreasing) orbital velocity, which makes the inner and outer loop diverge more (less) from eachother.
For the vertical cases, we find an opposite relation (in the metrics) as the inner loop has a different physical cause, i.e. suppression of the emission in the $\bm{k} \times \bm{B} \sim 0$ region.
This is also explained by the varying strength of relativistic beaming for the different orbit velocities, but it has the opposite effect with respect to the other magnetic field configurations, i.e. decreasing (increasing) inner loop size for decreasing (increasing) orbital speed.
This indicates that $\bm{k}$, and subsequently $\bm{k} \times \bm{B}$, are, unsurprisingly, quite sensitive to the orbital velocity.
The increase in $\bar{R}$ indicates that the $\bm{k} \times \bm{B} \sim 0$ region grows in size for increasing orbital velocity.

\paragraph{Plasma parameters:} 
Figure \ref{fig:QULP_other_i20} displays an exploratory model where we half our base electron density $n_e$ and double both the electron temperature $\Theta_e$ and magnetic field strength $B$.
Unsurprisingly, the (Stokes) quantities have changed with respect to the base model (shown in the top row of the figure).
The most notable difference are in LP and total \qu surface $S$ of the system, which  have both decreased substantially. The overall shape of the \qu structure remains relatively unaffected.
Variations in $\bar{S}$ and $\bar{R}$ are consistent with what was found in earlier figures.
Stokes $\mathcal{I}$ has almost quadrupled but Stokes $\mathcal{Q}$ and $\mathcal{U}$ have only doubled/tripled which explains both the decrease in LP and \quloop metrics.
Additionally, the intensity-averaged Faraday depth $\tau_V$ is three times smaller, but this is largely inconsequential as both cases are still very Faraday thin at $\tau_V \sim 10^{-2} \, \textrm{--} \, 10^{-3}$.
In the exploratory model, we have entered a relatively less polarized regime (at 230 GHz).
This example is illustrative of the fact that (polarized) synchrotron emission is highly sensitive to the underlying plasma quantities and is affected differently per magnetic field configuration.
This illustrates that fitting these models to observations would therefore be a non-trivial multi-parameter endeavour.

\subsubsection{Linear and circular polarization of bright spots} \label{res:linear_and_circular_polarization_lightcurves}
Bright spots orbiting in the disk are periodically highlighting (and adding to) emission regions with different linear and circular polarization. Can we use LP and CP light curves to infer the magnetic field topology or remove degeneracy between e.g. radial/toroidal cases (since they look similar in the EVPA evolution)?
The interpretation of LP and CP evolution is challenging because the polarization of light curves is modified by the beam depolarization. Also LP and CP magnitudes depend strongly on the underlying plasma parameters. 
However, the models we discuss in this work are all Faraday-thin which implies that almost all CP emission is intrinsic and Faraday conversion/rotation plays a negligible role in shaping \cp maps. This helps to understand the behaviour of the CP light curves.

In Figure~\ref{fig:spot_i90_a0} Stokes $\mathcal{I}$ exhibits variations that are expected for edge-on viewing angle. During the spot's passage the double peaked light curve is produced by the Einstein ring and followed by emission from Doppler beaming of the spot along {\bf its} orbit, as was already explained. 
The LP is mostly constant in all cases, except during the Einstein ring phase where LP decreases and the Doppler beamed phase where LP increases -- clearly seen in radial case, but is somewhat spin-dependent in toroidal case. 
The decrease in LP is caused by cancellation effect (beam depolarization) because places with opposite polarization direction cancel out. 
The LP increase is caused by amplification of the spot emission which is small in size and therefore amplifies only one polarization direction.
Due to symmetry, for the edge-on inclination with a vertical magnetic field geometry, CP becomes zero independently of the black hole spin.
For toroidal case CP decreases during Einstein ring moment, due to a cancellation effect, and then increases during the Doppler beaming time.
For radial case we see that CP flips sign during the spot's passage.
When the Einstein ring forms, CP is most negative and then gains positive sign during Doppler beaming which is a direct result of the underlying field topology.

In Figure~\ref{fig:spot_i20_a0} Stokes $\mathcal{I}$ exhibits small variations that are expected for low viewing angles. 
For the radial/toroidal cases we find a slightly enhanced flux when the spot is approaching the observer, but for vertical case the total intensity decreases due to suppression of synchrotron emissivity at $\zeta \approx 45^\circ$ (or inner \quloop passage). For radial fields the increase in Stokes $\mathcal{I}$, when spot is approaching the observer, is accompanied with a single peak in LP. For toroidal fields, increase in Stokes {\bf $\mathcal{I}$} is associated with two peaks in LP. 
Finally, for vertical fields, the decrease in Stokes I in the middle of spot's approach, is associated with a single peak in LP. 
When the spot is receding from the observer, LP increases to the maximum for the vertical case.
For vertical field case, the background flow polarity \cp is uniform across the image and the bright spot added to the model simply amplifies the CP, except during $\zeta \approx 45^\circ$ time moment (the inner \quloop time moment) when the CP light curves tend to show a small decrease. For toroidal fields the left and right parts of the \cp map have opposite polarities which results in a decrease of the CP amplitude when the spot is Doppler boosted. Finally, the radial field case is more complex because the top part of the \cp map has opposite polarity compared to the bottom one. Here one observes the decrease in CP light curves just before the Doppler boost moment marked with a hump in Stokes $\mathcal{I}$.

Let us conclude with a number of statements about CP and how one could utilize it to differentiate between the underlying magnetic field topologies during a \quloop event (with utilization of Figs. \ref{fig:spot_i20_a0} and \ref{fig:QUschematic}). Notice that our models assume that the spot contributes significantly to both LP and CP which, in reality, may not be always the case.
Starting with the vertical topology, we immediately find that the \quloop structure has a significant offset in \qu space with respect to the other magnetic field cases. 
The \quloop signature alone would therefore be sufficient to identify the vertical magnetic field (at this inclination and orbital velocity, see Fig. \ref{fig:QULP_keplerian_i20}). 
Additionally, during the inner \quloop passage, a slight drop in CP is perceivable.
Although the \quloops of the radial and toroidal cases span a similar \qu range, they have different CP behavior (and \quloop structure).
Due to the underlying \cp morphology we only see an increase in CP when the spot is approaching the observer (and the outer \qu loop is traversed) for the toroidal case.
For radial, we find the most complex \cp (image) morphology with an overall positive \cp structure with a section at $\zeta = 0^\circ \pm 25^\circ$ where \cp is negative.
This implies that when the inner \quloop is traversed (as the spot recedes from the observer) the CP keeps decreasing over time.
Note that the CP minimum is not reached when inner \quloop is still begin traversed, but rather when it has just entered the outer \quloop after which CP quickly increases.

\section{Discussion}\label{sec:discussion}
In this work we have studied polarimetric signatures of bright spots produced in accretion flow close to the black hole event horizon. 
Semi-analytic modelling 
enables the exploration of a parameter space without needing to run computationally expensive plasma simulations (GRMHD or GRPIC) where
plasma simulation's resolution may impact the results. Though not as complex as a full plasma simulation, our simplified approach allows for a relatively straight-forward decomposition of the various physical effects that are important in the system.

Even in the semi-analytic models the parameter space is very large so we can still only present a limited number of models. 
All models studied here have small optical depths $\tau_\nu \ll 1$ and small Faraday depths $\tau_F \ll 1$ (see Appendix~\ref{app:optical_depth_faraday_effects} for detailed definitions of optical and Faraday depths and their values for all models). 
We also fixed the magnetic field polarities and the orbital direction of the spot.
In Appendix~\ref{app:optical_depth_faraday_effects} we discuss the impact of the magnetic fields polarity, Faraday effects and spot rotation direction onto all polarimetric observables of a bright spot. 

Several works have presented polarized emission from hot spots around Kerr black hole.
\citet{broderick05} is one of the pioneering works on the semi-analytical modeling of spots polarization. 
They use an `optically thick sphere' to investigate the effect of a curved spacetime on polarized emission in a single vertically-aligned magnetic field configuration.
Especially the gravitational magnifying effect and the image centroid motion are topics they comment on.
In \citet{broderick06}, they include a disk background and change the magnetic field to toroidal.
Additionally, they improve their (optically thin) hot spot description by including a localized population of non-thermal electrons.
Much of our work is inspired by \citet{broderick05,broderick06}, but there are numerous differences - mostly with respect to our focus which lies predominantly on the polarimetric signatures.
While the aforementioned works mainly display net polarization fractions, we focus on the entire structure of the Stokes parameters (especially in \qu diagrams). 
Overall, when directly comparing the net polarization, there is little similarity between our model and the aforementioned ones by \citet{broderick05,broderick06}.

A more recent study by \citet{gelles21} outlines a toy model for synchrotron point source emission by analytically computing Kerr spacetime null-geodesics \citep{Gralla2020} and enforcing conservation of the Penrose-Walker constant, without any more sophisticated radiative transfer considerations.
This approach is fundamentally different from the numerical radiative transfer approach used in {\tt ipole} (while it is also a Penrose-Walker constant conserving scheme). 
The authors comment extensively on the \quloop structure they find.
However, they only model a hot spot without a disk background, which makes their result better comparable with NIR emission observed during flares in Sgr~A* than with observations at millimeter wavelengths.
Additionally, they effectively adopt a `fast light' approach where the spot is stationary when the geodesics are evaluated.
In our work, the spot continues its orbit while the radiative transfer equations are being integrated along the null geodesics. Lastly, note that they display \quloops and not the normalized \quloops we show in this work, so they are not intuitively comparable.
Their methodology does not produce images, which makes tracking down underlying effects difficult.
Even though our magnetic field descriptions are not identical (and the disk background is missing), we find very consistent \quloops structures in the low Faraday depth regime. The direct comparison of our model with \quloops from \citet{gelles21} is shown in \citealt{wielgus22polar}.

Another aspect investigated in this work, which was not addressed in detail previously, is the potential of using the circular polarization (or Stokes $\mathcal{V}$) to further constrain the underlying magnetic field structure.
If it is difficult to identify the model unambiguously based on the \quloop behavior alone, then it is possible to get additional information from the CP signature as we discussed in Sect. \ref{res:linear_and_circular_polarization_lightcurves}. 
A simple measurement of the CP sign can already help break this ambiguity, although imaging circular polarization structure (as shown in, e.g., Figure~\ref{fig:spot_i20_a0}) would be even more effective.
Note that polarity of the magnetic field is arbitrarily chosen and can possibly be reversed in reality, but the behavior of CP, especially while traversing an inner \quloop, will remain diverging (even when inverted) for the magnetic field cases evaluated in this work.

In a parallel work we demonstrated a remarkable consistency between predictions of our model and the linear polarization light curves of Sgr~A* observed with ALMA at 230 GHz shortly after an X-ray flare \citep{wielgus22polar}. Our findings strongly suggest low inclination of the Sgr~A* system and dominance of the vertical magnetic field component.
In the future, one can expand the semi-analytic hot spot model in a number of ways which include, but are not limited to, elliptical and/or non-equatorial orbits, separate magnetic field description for the spot, mixed (global) magnetic field descriptions, cooling of the hot spot over time, and presence of a non-thermal electron population. These more sophisticated models may provide a better fit to the existing data, and a deeper understanding of the physical processes. 

Our current base model consists of a RIAF disk and a hot spot, but one would expect an additional contribution of the jet \citep{blandford19}.
If one was to consider a disk-jet model \citep[e.g.,][]{Vincent2019} instead of a pure disk model, the emission from the disk would need to be reduced for the full model to end up at the same flux density.
As this will require rebalancing the plasma parameters (mainly through the electron temperature and density), it is possible that the disk will enter a more optically and/or Faraday thick regime.
Then, the background flow can become an internal Faraday screen, which may severely alter the polarimetric characteristics of the models \citep{moscibrodzka17,Jimenez2018}.
Additionally, it has been shown that strong variations in rotation measure as a result of the internal Faraday screen are expected from GRMHD simulations \citep{Ricarte2020}.

The physical origin and lifetime of hot spots are also uncertain. Alternatively to the plasmoid interpretation, the orbiting hot spots reported in Sgr~A* by \citet{gravity_plasmoids_18} and \citet{wielgus22polar} may correspond to flux tubes powered by reconnection and filled with hot electrons confined by the vertical magnetic field. Such features have been observed in numerical GRMHD simulations of magnetically arrested accretion disks \citep{ripperda22}. 
This origin would also support the notion that the hot spot emission should be (at least partially) described by non-thermal electron population, which is especially needed to explain the NIR observations \citep{chael17,davelaar18,davelaar19,fromm22}. The persistence and integrity of an orbiting hot spot also remain largely unknown. We expect hot spots to be disrupted by effects such as shearing in a differentially rotating flow \citep{tiede20} or by the Rayleigh-Taylor instability \citep{ripperda22} on dynamical timescales. Accounting for these effects would further modify the outcome of our calculations.

These topics are best investigated in a dedicated GRMHD and/or GRPIC simulations.
A number of studies have been conducted to better understand the behaviour and evolution of plasmoids, but these typically do not include a study of the radiation.
These simulation can be run within a shearing box \citep[see, e.g., ][]{murphy13,cerutti14a,sironi16} or global GRMHD and GRPIC simulations of magnetized accretion disks around a SMBH.
There, one generally finds more erratic plasmoid chains due to the less confining nature of the setup \citep{crinquand21,bransgrove21,ripperda20,ripperda22}.

\section{Summary}\label{sec:summary}

This study has outlined the various distinguishing polarimetric signatures of a circularly orbiting bright spot on top of a disk background at a frequency of 230 GHz.
For this model we have showcased the utility of \quloops and CP for establishing underlying magnetic field structure. 
We have shown what signatures are indicative of the underlying \emph{radial}, \emph{toroidal}, and \emph{vertical} magnetic fields. In the following paragraphs, we summarise the main findings of this work.

\begin{enumerate}[(i)]
\item The shape and size of a \quloop is indicative for, at least, three parameters we investigate, namely; inclination $i$, magnetic field, and BH spin ($a_*$).
The inner \quloop structure is particularly informative. 
The point where the \qu path over time moves from the outer to inner loop was previously referred to as the crossing-point. 
This point lies on opposite sides for the \emph{toroidal}/\emph{vertical} magnetic field cases and the \emph{radial} case. This does not change with inclination (including $i > 90^\circ$), but the exact position of the crossing point is sensitive to rotation of the image. 
Additionally, as becomes clear from Figure~\ref{fig:QUschematic}, the timing (of entering/exiting of the inner loop) is strikingly different between the \emph{radial}/\emph{toroidal} and \emph{vertical} cases.
While for the former case the division between inner and outer \quloop lies in the approaching versus receding part of the orbit, for the \emph{vertical} case, it is connected to the part of the orbit where the direct emission from the spot disappears (as outlined in Sect. \ref{res:variable_emission_polarization}).

\item Moreover, inclination ($i$) is especially influential on the \quloop structure for the \emph{vertical} magnetic field case, as demonstrated in Figs. \ref{fig:QU_spot_disk} and \ref{fig:QULP_incl}. In this configuration the inner loop structure disappears for $i \gtrsim 25^\circ$ and for even greater inclination angles one only recovers thin crescents which have little resemblance to the typical \quloop structure.
Another influential parameter is the orbital velocity of the system, particularly for the \emph{vertical} case (as highlighted in Fig. \ref{fig:QULP_keplerian_i20}).
To a lesser degree, this behavior is also present for \emph{radial} and \emph{toroidal} cases as the inner loop relative size increases as function of inclination.
However, for those cases, we find that even for near edge-on viewing angles there is still some inner loop structure present as shown in Fig.~\ref{fig:QU_spot_disk}.

\item The \emph{vertical} magnetic field case behaves substantially different from the two others -- both in terms of \quloop structure and LP strength. 
Overall, the \emph{vertical} cases are more ${\mathcal Q}$-negative and variation of certain parameters results in significant shift in Stokes ${\mathcal Q}$ (as shown in Figs. \ref{fig:QULP_incl}, \ref{fig:QULP_i20}, and \ref{fig:QULP_keplerian_i20} for $i$, $a_\ast$, and $v_\textrm{kep}$, respectively).
The \emph{radial} and \emph{toroidal} cases generally span an almost identical \qu range and display only minor variations in Stokes ${\mathcal Q}$.
Interestingly, Stokes ${\mathcal U}$ is (almost) insensitive to any of the parameter variations with an average value of $\mathcal{U}/\mathcal{I} \approx 0$ for the \emph{radial}/\emph{toroidal} cases.
Some variation in $\mathcal{U}$ is found for the vertical case (where it becomes more positive), but this remains minimal when compared to the variations in $\mathcal{Q}$.

\item When \qu structure is similar or unclear, CP could break the degeneracy. 
\emph{Radial} and \emph{toroidal} cases for low inclinations ($i \lesssim 30\degr$) would observationally be hard to differentiate. 
However, they do differ structurally in CP.
The region close to the photon ring has negative CP (for all cases) while the surrounding emission structure (from radii beyond ISCO) has a predominantly positive CP for the \emph{radial} case and negative CP for the \emph{toroidal} case for $40\degr \gtrsim i \gtrsim 150\degr$. 
However, for edge-on viewing angles ($70\degr \lesssim i \lesssim 110\degr$), CP structure becomes more similar, but then one can differentiate based on LP.
Therefore, we argue that all Stokes parameters, $\mathcal{I}$, $\mathcal{Q}$, $\mathcal{U}$, and $\mathcal{V}$, are required to robustly determine dominant underlying magnetic field structure.

\end{enumerate}

\section*{Acknowledgements}
We thank Alejandra Jim\'enez-Rosales, Hector Olivares, Aristomenis Yfantis, and Zachary Gelles for the insightful discussions and comments that have improved the quality of the manuscript. 
Additionally, we thank the internal EHT reviewer Michi Baubock for his comments.
MM acknowledges support by the Dutch Research Council (NWO) grant No. OCENW.KLEIN.113.
MW thanks Alexandra Elbakyan for her contributions to the open science initiative.


\bibliographystyle{aa}
\bibliography{references}



\appendix

\section{Linear and circular polarization} \label{app:lin_cir_polarization}

\begin{figure}
    \centering
    \includegraphics[width=0.48\textwidth]{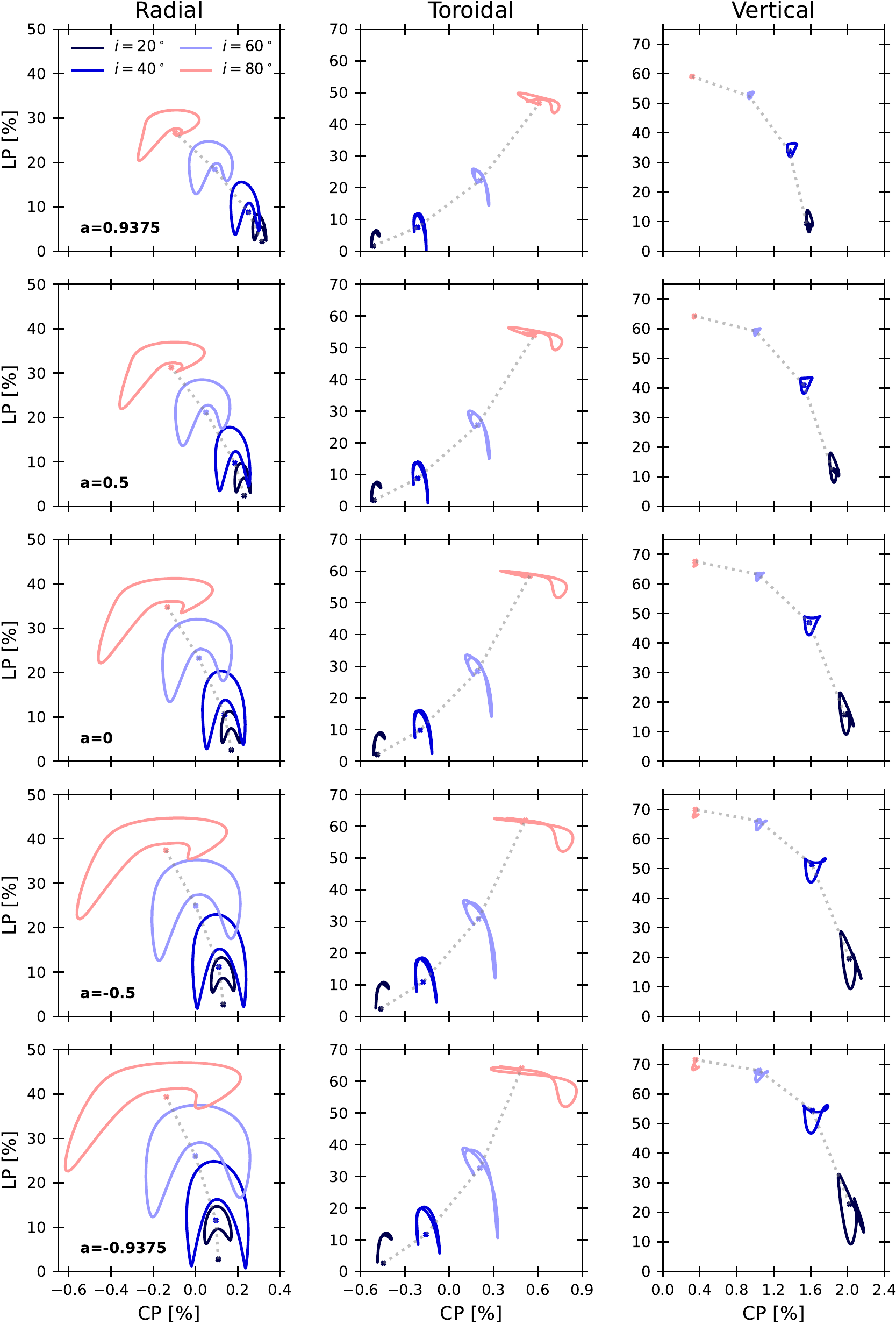}
    \caption{LP versus CP diagrams for models with different magnetic fields, black hole spins and viewing angles. The crosses denote the disk-only values. The dotted lines are shown to highlight a trend.}
    \label{fig:CPLPdef}
\end{figure}

Figure~\ref{fig:CPLPdef} displays the relation between LP and CP for all spins and inclinations of $i=20^\circ, \, 40^\circ, \, 60^\circ, \, 80^\circ$.
Each magnetic field configuration displays a unique relation between LP and CP.
For increasing inclination, LP rises for all cases.
However, for increasing inclination, we find that CP decreases radial and vertical cases and increases for the toroidal case.
Another diverging feature is the size of path followed in the LP and CP. 
The radial case is more variable in both LP and, especially, CP than the other cases. Additionally, it has a clear crescent-like shape, which is only deformed for the higher inclination cases (due to boosting).
The toroidal case has a vaguely similar shape, but it has almost no difference between the part of the orbit when the spot is receding (lower part of CPLP structure) and when the spot is approaching (upper part of CPLP structure).
Lastly, for the vertical case, we observe that there is little CP and LP variation during the spot's orbit for any inclination.
Overall, for all cases, we find that CP and LP variability decreases with increasing spin, which is due to the increased contribution of the spot to the overall flux density for lower spin cases.

\begin{figure}
    \centering
    \includegraphics[width=0.48\textwidth]{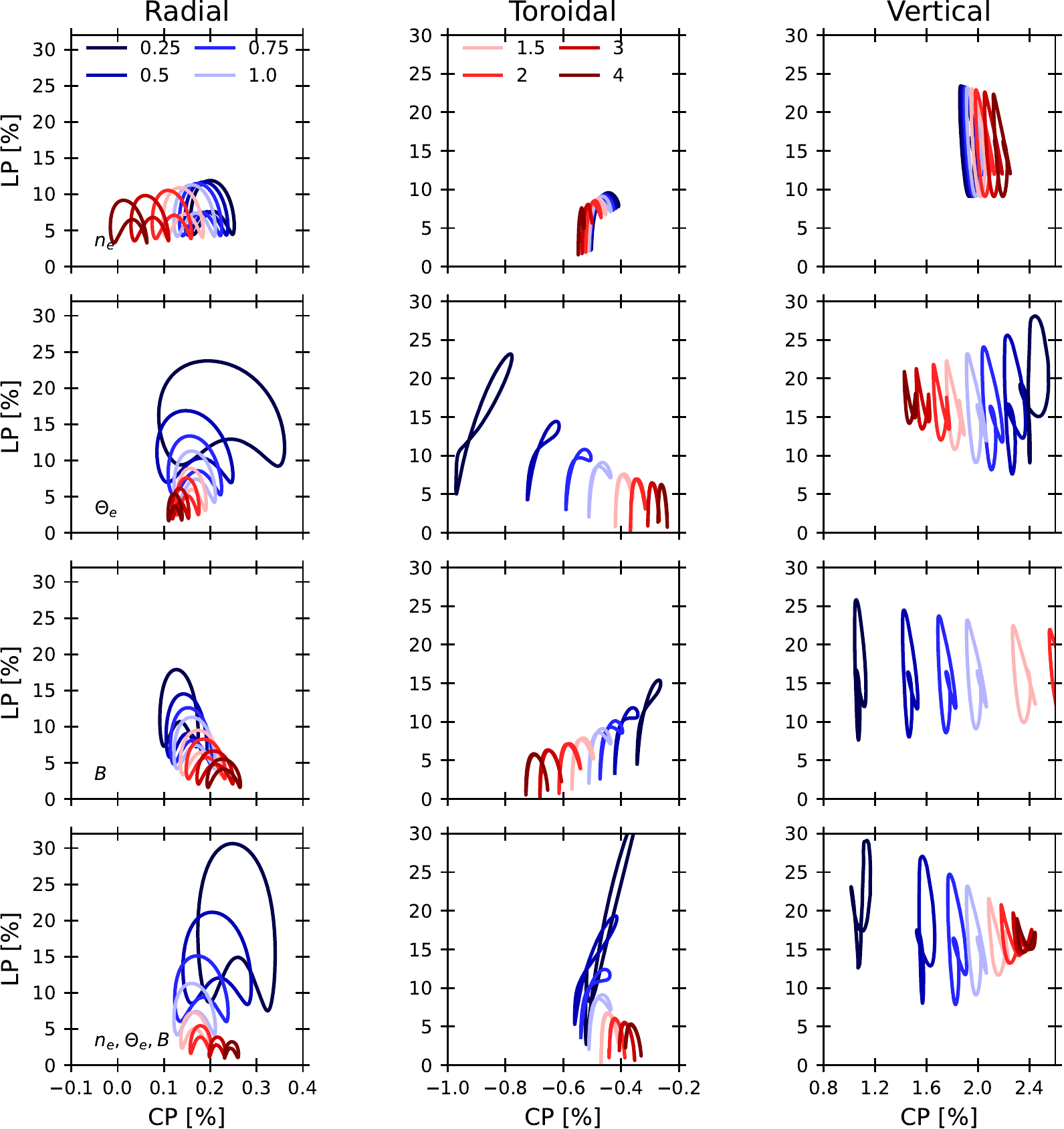}
    \caption{LP versus CP diagrams for models with different magnetic fields, black hole spin of $a_*=0$, and viewing angle of $i=20^\circ$. In each row we show the base plasma property (i.e., $n_e$, $\Theta_e$, $B$) which are color-coded according to the multiplication factor with which they are scaled. The bottom row shows the behavior where all quantities ($n_e$, $\Theta_e$, $B$) are scaled simultaneously.
    }
    \label{fig:CPLPplasmatests}
\end{figure}

Figure \ref{fig:CPLPplasmatests} continues to outline the relation between LP and CP, but here we vary the underlying plasma parameters modestly to investigate slight deviations from our base configuration.
The base model has $a_*=0$ and $i=20^\circ$ and corresponds to the modification factor of 1 in the figure.
In the top row, we vary only the electron (number) density $n_e$.
There, we find that LP is almost invariant for variations in $n_e$.
The LP-CP structure decreases slightly and CP becomes more negative as the $n_e$ modification factor increases.
This effect is largest for the radial case, but is also present for the toroidal case.
For the vertical case, we find that the relation is reversed - CP increases minimally with increasing multiplication factor.
In the second row, we vary electron temperature $\Theta_e$.
There, we find a larger response both in LP-CP structure alterations and corresponding CP values.
For the radial case, we observe that both the LP and CP variation decreases if the temperature increases.
For the toroidal case, CP is most negative and largest in size for smallest $\Theta_e$ value but increases when the multiplication factor increases.
A similar relation is observed for the vertical case, but again in reverse - CP becomes smaller for increasing multiplication factor.
In the third row, we observe that variations in magnetic field strength $B$ once again produce unique alterations to the model's LP and CP.
For the radial case, we find that LP decreases and CP increases for increasing $B$.
For the toroidal case, both LP and CP increase as $B$ increase.
For the vertical case, LP remain constant but we observe a significant increase in CP.
For the bottom row, all quantities ($n_e$, $\Theta_e$, $B$) are scaled simultaneous. 
The behaviour here clearly shows that trying to predict the plasma behaviour based on slight deviations to the base model is already a non-trivial endeavour.
Each model once again describes a relation between LP and CP that's different from the behaviour in the three top rows. Interestingly, both the radial and toroidal cases react similarly, but the vertical case is most similar to the $B$ variation.

\section{Circular polarization from regions close to the photon ring}\label{app:photon_ring}
\begin{figure}
    \centering
    \includegraphics[width=.48\textwidth]{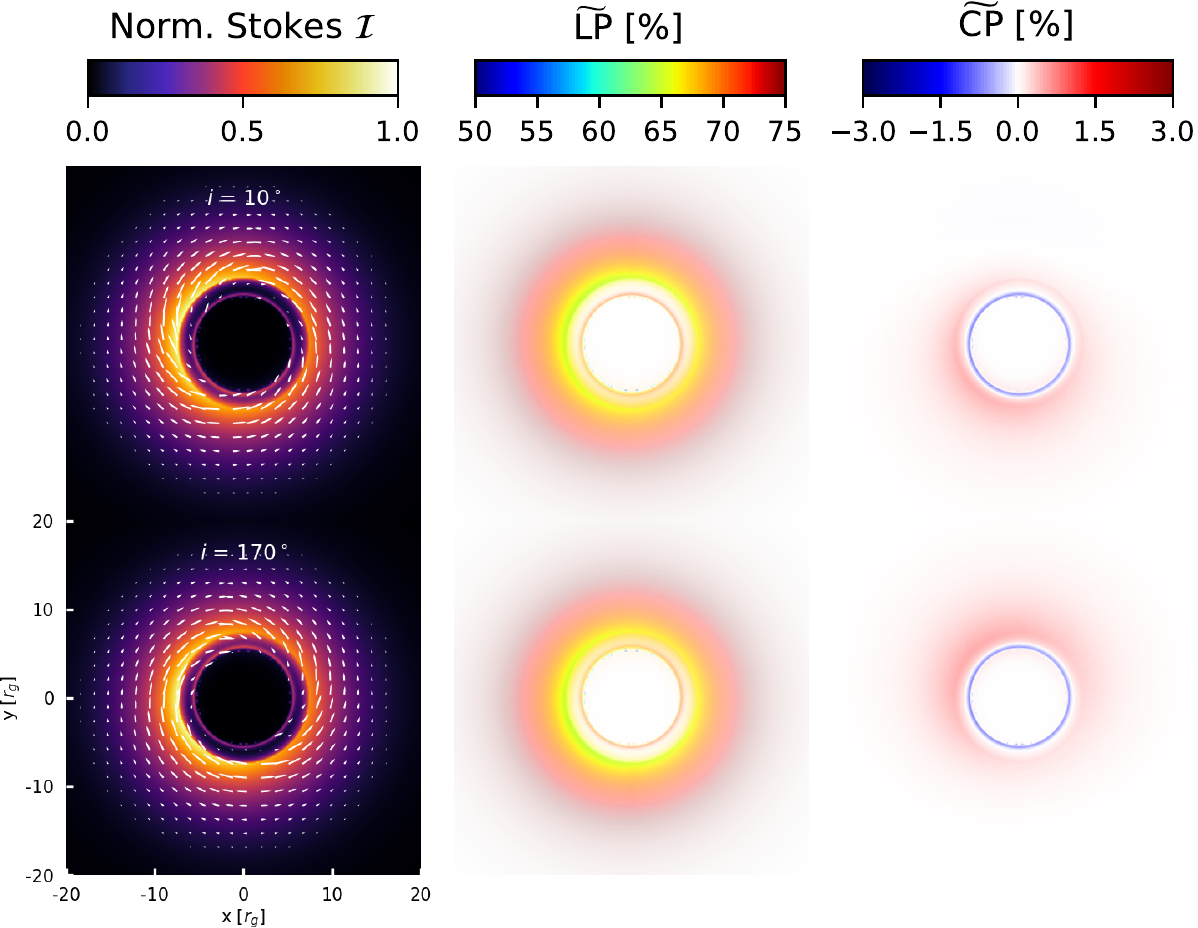}
    \caption{
    Stokes $\mathcal{I}$ (\emph{left} panel) and linear (\emph{middle} panel) and circular (\emph{right} panel) polarization images of the background accretion flow for observer viewing angles of $i = 10^\circ$ and $i=170^\circ$ and a \emph{radial} magnetic field topology. 
    }
    \label{fig:photonring_radial}
\end{figure}

\begin{figure}
    \centering
    \includegraphics[width=.48\textwidth]{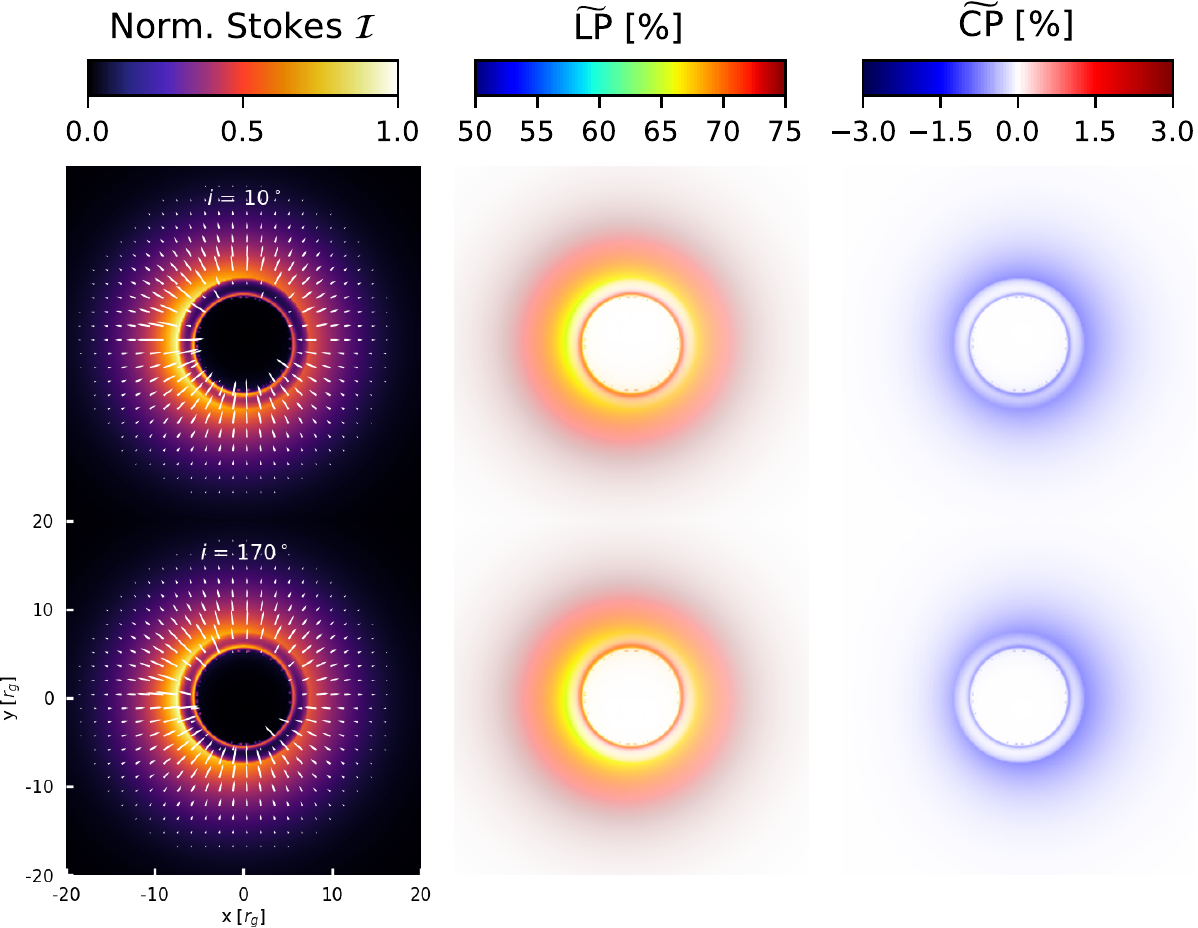}
    \caption{Same description as for Fig. \ref{fig:photonring_radial} only for the \emph{toroidal} magnetic field topology. }
    \label{fig:photonring_toriodal}
\end{figure}

A feature that further magnifies the differences between the magnetic field cases is the \cp signature of the lensed emission directly outside the photon ring.
The photon ring is a infinitesimal feature, with a well-defined radius \citep[see, e.g.,][]{bardeen72}, that is not well-resolved in our images but generally has a similar polarization (if slightly reduced) with respect to the lensed emission directly outside it \citep{jimenez21}.
For the rest of the section, we will be discussing the emission that lies directly outside the photon ring that we will simply refer to as "emission" from now on.
For the radial case, we find a strictly negative \cp, which is most pronounced for face-on viewing angles ($110^\circ \lesssim i \lesssim 70^\circ$) as demonstrated in Figs. \ref{fig:disk_incl_a0} and \ref{fig:photonring_radial}.
For the toroidal case, we find the vertical division as outlined in Sect. \ref{res:linear_and_circular_polarization_lightcurves} persists in the photon ring's \cp signature (as shown in Figs.~\ref{fig:disk_i90_a0} and \ref{fig:disk_incl_a0}) - positive \cp on the left and negative \cp on the right. For the face-on viewing angle (from top or bottow), we find no apparent difference in the sign of \cp between photon ring and disk as shown in Fig.~\ref{fig:photonring_toriodal}.

\begin{figure}
    \centering
    \includegraphics[width=.48\textwidth]{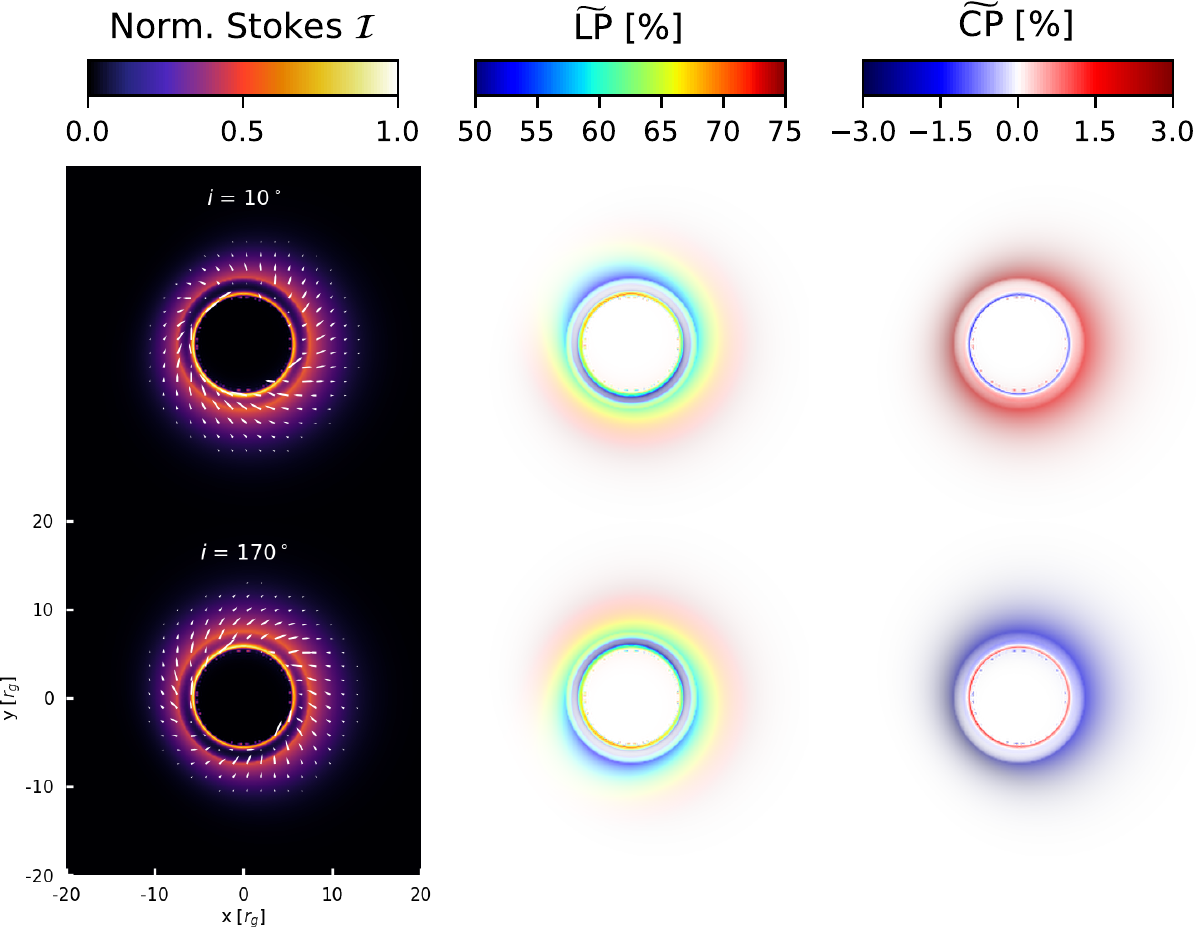}
    \caption{Same description as for Fig. \ref{fig:photonring_radial} only for the \emph{vertical} magnetic field topology. }
    \label{fig:photonring_vertical}
\end{figure}

The vertical case is deviating in the sense that the \cp signature of the emission directly outside the photon ring is strongly determined by viewing angle.
For $i < 90^\circ$, we find negative \cp for the photon ring and an overall positive \cp from disk and spot emission. For $i>90^\circ$, the situation is reversed.
This is clearly shown in Fig.~\ref{fig:photonring_vertical}.
As can be seen in Fig. \ref{fig:disk_i90_a0}, the edge-on view reveals two modes - a top half which has negative \cp and a bottom half with positive \cp for the photon ring. This overall \cp structure is oppositely polarized with respect to the photon ring.
The dividing line, which lies along the equatorial plane, is unpolarized and splits these regions.

If we were to look at \lp alone, then the radial and toroidal case describe a similar structure which would be hard to differentiate.
For \cp, however, the structure is completely different, which allows for a straight-forward identification.

\section{Optical/Faraday Depth and other effects} \label{app:optical_depth_faraday_effects}

\begin{figure*}
    \centering
    \includegraphics[width=0.49\textwidth]{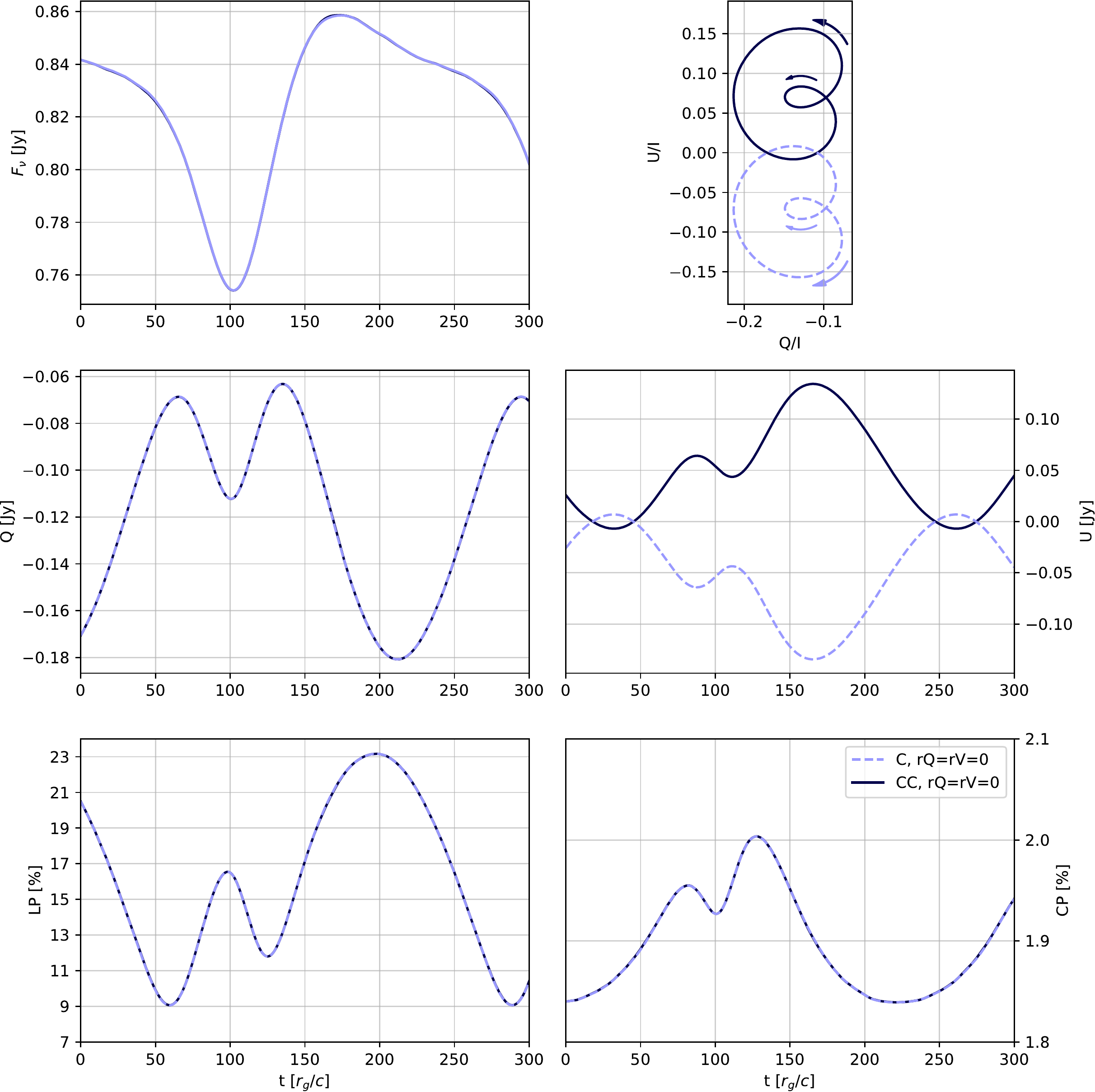}
    \includegraphics[width=0.49\textwidth]{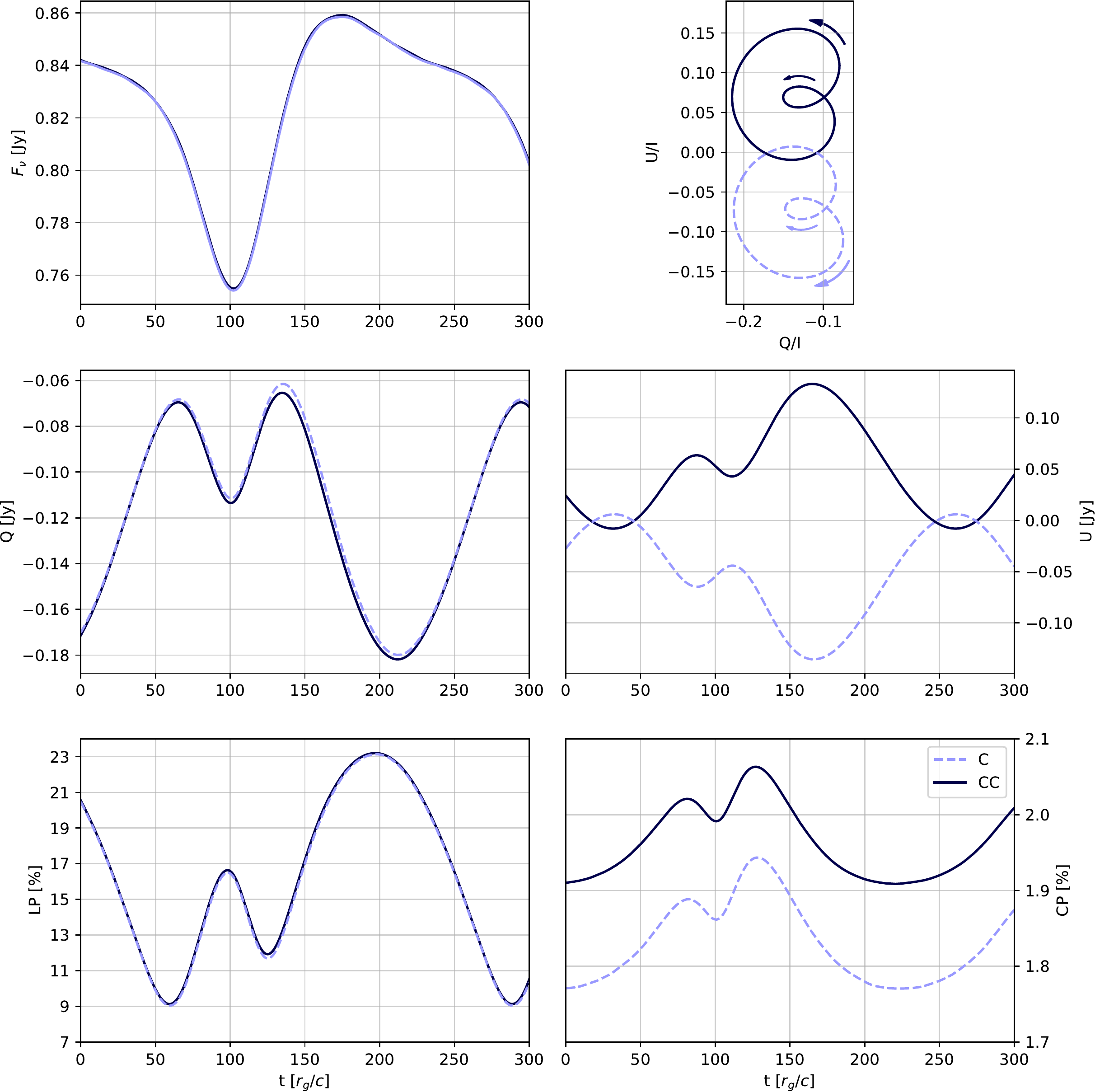}
    \caption{Impact of Faraday effects on the Stokes $\mathcal{IQUV}$, LP and CP evolution in model with vertical magnetic field viewed at $i=20 \deg$. In the left panel we show light curves of a model where we disabled the Faraday rotation and conversion effects by setting the Faraday rotativities to zero (radiative transfer model with $\rho_{QV}=0$). In the right panels we show the same model but with the Faraday effects taken into account. Solid/dashed lines in each sub-panel indicate model with accretion disk and the bright spot on clockwise (C) / counterclockwise (CC) orbit. Arrows in the QU panels indicate the direction of time evolution. 
    }
    \label{fig:app_faraday_optical_depth}
\end{figure*}

Models considered in this work are mostly optically thin, with optical depth of $\tau_a \sim 10^{-3} - 10^{-2}$.
Models are also Faraday thin, with Faraday depths $\tau_V \sim 10^{-3} - 10^{-2}$. The Faraday thickness defined as $\tau_V=\int \rho_V dl$ (where $\rho_V$ is the transfer coefficient for Faraday rotation) gives us an estimate of how many rotations (in radians) the EVPA is undergoing when a ray of light is transported from the emission origin to the observer.
This means that Faraday rotation has small impact on linear polarization of models. 

Nevertheless, we discuss in more detail how Faraday effects impact the linear and circular polarization of our models. This is done for two reasons: i) Faraday conversion can alter circular polarization, ii) Faraday effects may break degeneracies in observables such as fractional linear and circular polarization of hot spot models rotating in different directions (on clockwise and counter-clockwise orbits) or in models with opposite magnetic field polarity. Even for completely Faraday thin models changing the magnetic field polarity results in Stokes ${\mathcal U}$ with opposite sign. 

The discussion is based on analysis of a model of accretion disk with a spot around non-rotating black hole threaded by vertical magnetic field seen at a viewing angle of 20 degrees. In this model a QU loop is formed and circular polarization experiences significant variation.  

First we consider a case when both Faraday rotation and Faraday conversion are absent ($\rho_{Q,V}=0$). Changing the direction of system rotation from clockwise to counterclockwise reverses the sign of Stokes ${\mathcal U}$ and hence the direction in which the QU loop is traversed (the clockwise/counterclockwise rotating systems with display clockwise/counterclockwise QU loops). The fractional linear polarization and circular polarization remain intact. In the model without Faraday effects the circular polarization is produced intrinsically, hence the change of the system rotation changes neither the sign nor the magnitude of the circular polarization. 
In Fig.~\ref{fig:app_faraday_optical_depth} (left panel) we demonstrate that this is true for our chosen model, as expected.
Notice also that, for optically thin emission, when Faraday conversion is absent, the sign of the Stokes ${\mathcal V}$ depends only on the polarity of the magnetic fields. Changing the magnetic field polarity to the opposite one will simply reverse the sign of ${\mathcal V}$. 

Secondly, we consider the same model but with the Faraday effects included (shown in the right panels in Fig.~\ref{fig:app_faraday_optical_depth}). Stokes ${\mathcal QU}$ are slightly rotated by Faraday rotation, hence clockwise and counter-clockwise rotating models will have slightly different linear polarizations and the clockwise system QU loop will not be identical reflection of the counterclockwise rotating system. The LP will be changed. It is interesting to note that reversing the sign of the magnetic fields in this case will result in Faraday rotation in the opposite direction (since $\rho_V \sim B_{||}$ along the line of sight), so the QU loop of clockwise model with $+B$ should be a reflection of the QU loop of the counterclockwise model with $-B$ (this should be true only for spin zero). 
Notice also that when Faraday effects are included, the magnitude of Stokes ${\mathcal V}$ is changed meaning that in our models significant portion of the observed circular polarization is produced in the process of Faraday conversion. 

\label{lastpage}
\end{document}